\documentclass[prl,superscriptaddress,longbibliography,twocolumn]{revtex4-1}
\usepackage{geometry}
\usepackage{amsmath}
\usepackage{amssymb}
\usepackage{graphicx}
\usepackage{hyperref}
\usepackage{braket}
\usepackage{bm}
\usepackage[usenames,dvipsnames]{color}

\graphicspath{./}

\newcommand{\be}{\begin{equation}}
\newcommand{\ee}{\end{equation}}

\newcommand{\bQ}{{{\bf{Q}}}}
\newcommand{\br}{{{\bf{r}}}}

\newcommand{\bq}{{\bf{q}}}

\newcommand{\bea}{\begin{eqnarray}}
\newcommand{\eea}{\end{eqnarray}}
\newcommand{\beal}{\begin{align}}
\newcommand{\eeal}{\end{align}}
\newcommand{\ra}{\rangle}
\newcommand{\la}{\langle}

\newcommand{\upa}{\uparrow}
\newcommand{\dna}{\downarrow}

\newcommand{\fQ}{F${\cal Q}$} 
\newcommand{\nQ}{N${\cal Q}$}
\newcommand{\spQ}{Sp${\cal Q}$}
\newcommand{\fO}{F${\cal O}$}
\newcommand{\nO}{N${\cal O}$}
\newcommand{\afQ}{AF${\cal Q}$}

\begin{document}

\title{Two-stage multipolar ordering in Pr(TM)$_2$Al$_{20}$ Kondo materials}

\author{Frederic Freyer}
\affiliation{Institute for Theoretical Physics, University of Cologne, 50937 Cologne, Germany}
\author{Jan Attig}
\affiliation{Institute for Theoretical Physics, University of Cologne, 50937 Cologne, Germany}
\author{SungBin Lee}
\affiliation{Department of Physics, Korea Advanced Institute of Science and Technology, Daejeon, 34141, Korea}
\author{Arun Paramekanti}
\affiliation{Department of Physics, University of Toronto, Toronto,
  Ontario M5S 1A7, Canada}
  \author{Simon Trebst}
\affiliation{Institute for Theoretical Physics, University of Cologne, 50937 Cologne, Germany}
\author{Yong Baek Kim}
\affiliation{Department of Physics, University of Toronto, Toronto,
  Ontario M5S 1A7, Canada}

\begin{abstract}
Among heavy fermion materials, there is a set of rare-earth
intermetallics with non-Kramers Pr$^{3+}$ $4f^2$ moments which exhibit a rich phase diagram with intertwined quadrupolar
orders, superconductivity, and non-Fermi liquid behavior. However, more subtle broken symmetries such as 
multipolar orders in these Kondo materials remain poorly studied. Here, we argue that {\em multi-spin} interactions between
local moments beyond the conventional two-spin exchange must play an important role in Kondo materials near the
ordered to heavy Fermi liquid transition. We show that this
drives a plethora of phases with coexisting multipolar orders and multiple thermal phase transitions,
providing a natural framework for interpreting experiments on the Pr(TM)$_2$Al$_{20}$ class of compounds.
\end{abstract}

\maketitle


The celebrated Doniach picture of Kondo materials captures their evolution from magnetically ordered phases of local
moments to the eventual heavy Fermi liquid phase when the 
local moments get fully incorporated into the Fermi sea \cite{doniach1977kondo, ruderman1954indirect, kasuya1956theory, yosida1957magnetic}. These systems provide a fertile ground for the emergence of 
novel electronic and magnetic ground states as well as exotic quantum phase transitions \cite{stewart1984heavy,fisk1995physics,coleman2001fermi,gegenwart2008quantum,si2010heavy}.
While magnetic ordering of a periodic array of local moments
and its influence on Kondo physics has been studied extensively, 
subtler forms of broken symmetry such as multipolar orders remain less
explored \cite{morin1982magnetic,cox1987quadrupolar,cox1999exotic,kitagawa1996possible,caciuffo2003multipolar,suzuki2005quadrupolar,kuramoto2009multipole,lee2015optical}.
In this context, recent experiments on the rare-earth intermetallics
Pr(TM)$_2$Al$_{20}$ (TM=Ti,V) and PrIr$_2$Zn$_{20}$ are significant,
showing rich phase diagrams as a function of temperature, pressure, and 
magnetic field, exhibiting quadrupolar orders, non-Fermi liquid physics,
and superconductivity (SC)\cite{sakai2011kondo,koseki2011ultrasonic,sakai2012thermal,sato2012ferroquadrupolar,onimaru2016exotic,onimaru2011antiferroquadrupolar,sakai2011kondo,shimura2013evidence,onimaru2012simultaneous,onimaru2010superconductivity,sakai2012superconductivity,matsubayashi2012pressure,matsubayashi2014heavy,tsujimoto2014heavy,iwasa2017evidence,taniguchi2016nmr}.

In these systems, Pr$^{3+}$ ions have a non-Kramers ground state doublet,
which acts as a pseudospin-$1/2$ degree of freedom on the diamond lattice \cite{sato2012ferroquadrupolar,onimaru2016exotic}.
As explained later, two components of this pseudospin
carry a quadrupolar moment while the third component describes an
octupolar moment, so their ordering would respectively correspond to time-reversal-even
quadrupolar and time-reversal-odd octupolar symmetry breakings \cite{shiina1997magnetic}.
Such ordering is expected to be driven by a
Kondo-coupling to conduction electrons arising from
TM and Al in Pr(TM)$_2$Al$_{20}$ (TM = Ti, V).
Indeed, experiments suggest ferroquadrupolar (\fQ) ordering in PrTi$_2$Al$_{20}$ at $T_1 \approx 2$K,
well above the superconducting transition temperature $T_c \approx 0.2$K \cite{sakai2011kondo,sakai2012superconductivity,koseki2011ultrasonic,sato2012ferroquadrupolar,taniguchi2016nmr}.
A recent series of experiments on PrV$_2$Al$_{20}$ discovered two closely spaced
consecutive thermal transitions, at $T_1 \approx 0.8$K and  $T_2 \approx 0.7$K, again 
well above the superconducting $T_c\approx 50$mK, with evidence that the higher transition at $T_1$
is due to antiferroquadrupolar (\afQ) order \cite{sakai2011kondo,tsujimoto2014heavy,tsujimoto2015anomalous}.
Understanding such multipolar orders is important for clarifying the phase diagram 
of these
heavy fermion systems, including the origin of SC.

On general grounds, the Doniach phase diagram of heavy fermion materials suggests that the weak 
Kondo coupling regime would lead to local-moment order driven by RKKY interactions,
while the strong Kondo coupling regime would lead to a hybridized heavy Fermi liquid (FL) with a large Fermi surface (FS)\cite{ruderman1954indirect, kasuya1956theory, yosida1957magnetic,stewart1984heavy, lohneysen2007fermi}.
The transition between these phases might be driven by
increasing pressure or by choice of the TM ion; for instance, PrV$_2$Al$_{20}$ appears to have stronger hybridization than PrTi$_2$Al$_{20}$ \cite{sakai2011kondo}. 
While attention has been mainly focussed on the
quadrupolar orders in such rare-earth intermetallics, our main observation is that the broader class of ordered phases could also involve the 
octupolar degrees of freedom driven by higher order {\em multi-spin} interactions, which have not been carefully explored.

One route to understanding the origin of such multi\-spin interactions is to see that the `small' to `large' FS transition is driven by
increasing hybridization. This will lead to the importance of higher order RKKY interactions, which can involve more than two spins.
Alternatively, let us consider the Doniach phase diagram from the viewpoint of an orbital-selective
Mott transition of the local moments \cite{de2005orbital,de2009orbital}. In this case, the ordered phase with a small FS is an `ordered Mott insulator' of the local
moments, while the hybridized FL is a `metallic phase' of the local moments. 
In analogy with organic Mott insulators, where four-spin ring exchange interactions near the Mott transition have been proposed to drive a 
quantum spin liquid with a spinon Fermi surface \cite{Motrunich2005,Motrunich2007,sheng2008boson,sheng2009spin,grover2010weak},
we expect that upon approaching the `Mott insulator' to `metal' transition of the rare earth moments, similar multi-spin interactions might become
significant and
drive exotic phases of the local moments. This idea finds support in recent {\it ab initio} and phenomenological calculations on certain Kondo materials \cite{Staunton2017,akagi2012hidden}.

In this Letter, we consider a frustrated local-moment model with two-spin and four-spin interactions, that are allowed by 
symmetry associated with the local environment of Pr$^{3+}$ ions and their coupling 
to the conduction electrons.
Since our main interest is the interplay between different multipolar orders
and their thermal phase transitions, we employ mean field theory and Monte Carlo simulations
to investigate the thermal phase diagram of this model.
Our key result is that such interactions can lead to ground states with {\em coexisting} multipolar orders; we show that
this can lead to a single or two-stage multipolar thermal transitions, and present results on the effect of a 
magnetic field. We discuss how this provides a natural
framework to interpret the experiments on PrTi$_2$Al$_{20}$ and PrV$_2$Al$_{20}$,
which is thus also of potential importance for other heavy fermion materials.

{\em Model.---}
In Pr(TM)$_2$Al$_{20}$ (with TM=Ti, V), the $4f^2$ Pr$^{3+}$ ion lives in a $T_d$ local environment, arising from the Frank Kasper cage formed by
16 neighboring  Al ions \cite{onimaru2016exotic}. Inelastic neutron scattering and specific heat studies have shed light on the local spectrum of the Pr$^{3+}$ ion,
arising from crystal field splitting of the $J=4$ angular momentum multiplet
\cite{sakai2011kondo,sato2012ferroquadrupolar}. These
indicate a $\Gamma_3$ non-Kramers doublet ground state separated from the next $\Gamma_4$ triplet of states by an energy gap $\sim 50$K. At temperatures
$T \ll 50$K, we can effectively ignore these excited crystal field multiplets \cite{sakai2011kondo}.
Thus, for the low energy
physics of these materials, especially the broken symmetry phases found at $T \lesssim 5$K, it is sufficient to consider a 
model of conduction electrons Kondo-coupled to this $\Gamma_3$ doublet, whose
wavefunctions are \cite{sato2012ferroquadrupolar,hattori2014antiferro}
\bea
\ket{\Gamma_3^{(1)}} &=& \frac{1}{2} \sqrt{\frac{7}{6}} \ket{4} - \frac{1}{2}  \sqrt{\frac{5}{3}} \ket{0} + \frac{1}{2} \sqrt{\frac{7}{6}} \ket{-4} \, \nonumber \\
\ket{\Gamma_3^{(2)}} &=& \frac{1}{\sqrt{2}} \ket{2} + \frac{1}{\sqrt{2}} \ket{-2}.
\eea
Using these, we can define the pseudospin-$1/2$ basis
$\ket{\upa} \!\equiv \! \frac{1}{\sqrt{2}} (\ket{\Gamma_3^{(1)}} \!+\! i \ket{\Gamma_3^{(2)}})$ and
$\ket{\dna} \!\equiv\! \frac{1}{\sqrt{2}} (i \ket{\Gamma_3^{(1)}}\! +\!\ket{\Gamma_3^{(2)}})$.
We identify the corresponding pseudospin operators in terms of Stevens operators 
$O_{22} \!=\! \frac{\sqrt{3}}{2} (J^2_x \!-\! J^2_y)$, $O_{20} \!=\! \frac{1}{2}(3 J^2_z\!-\!J^2)$, and 
$T_{xyz} \!=\! \frac{\sqrt{15}}{6} \overline{J_x J_y J_z}$ (overline denoting a symmetrized product),
as $\tau_x \!=\! - \frac{1}{4} O_{22}$, $\tau_y \!=\! -\frac{1}{4} O_{20}$, and $\tau_z \!=\! \frac{1}{3\sqrt{5}} T_{xyz}$\cite{stevens1952matrix,lea1962raising}.
Here, $(\tau_{x},\tau_y) \!\equiv\! \vec \tau^\perp$
describes a time-reversal invariant quadrupolar moment, while $\tau_z$ describes a time-reversal odd octupolar 
moment. In addition, the point group symmetry of the Pr$^{3+}$ ion includes an $S_{4z}$ operation under which 
$\tau^\pm \to -\tau^\pm$, and a $C_{31}$ operation under which $\tau^\pm \to {\rm e}^{\pm i2\pi/3} \tau^\pm$.

With this in mind, we consider a symmetry-allowed model of short-distance two-spin exchange between the pseudospin-$1/2$  local moments 
$\vec \tau$, supplemented with the simplest four-spin interaction that couples quadrupolar and octupolar degrees of freedom,
\bea
H \!=\! \frac{1}{2} \sum_{i,j} \! J_{ij} (\vec \tau^\perp_i \cdot \vec \tau^\perp_j \!+\! \lambda \tau^z_i \tau^z_j) - K \!\!\!\!\!\!\!  \sum_{\la \la ij \ra \la km \ra \ra} \!\!\!\!\!\! \vec \tau^\perp_i \cdot \vec \tau^\perp_j \tau^z_k \tau^z_m \,. \quad
\label{eq:JKmodel}
\eea
We will assume $J_{ij} = J_1, J_2$ for nearest and next-nearest neighbors respectively, and ignore further neighbor two-spin interactions.
For the four-spin coupling, the notation 
$\la \la ij \ra \la km \ra \ra$ means that we consider a nearest-neighbor pair $\la i j \ra$ coupled to a distinct nearest-neighbor pair $\la km \ra$, such that
the two pairs are separated by a single bond, leading to the shortest four-site cluster
\footnote{We note in passing that there are symmetry allowed three-spin interactions, but they are highly frustrated; these will be explored elsewhere.}.

We consider the easy-plane regime, $\lambda\! < \! 1$, so that the two-spin interactions favor
quadrupolar $\tau^{x,y}$ order over octupolar $\tau^z$ order as is observed 
in many of these compounds. While $J_1 \!<\! 0$ will drive
\fQ~order, as observed in PrTi$_2$Al$_{20}$, increasing pressure might lead to \afQ~orders, 
either via a frustrating $J_2/|J_1| \!>\! 0$ which leads to incommensurate spiral order (\spQ), or via a sign change $J_1\! > \! 0$ which will lead to
commensurate N\'eel quadrupolar order (\nQ) \cite{sato2012ferroquadrupolar,sakai2011kondo}.
Our main insight is that while the two-spin interactions alone will favor pure quadrupolar order, 
four-spin interactions will generically lead to coexisting multipolar orders. For $K>0$, quadrupolar orders with nearest-neighbor 
$\la \vec \tau^\perp_i \cdot \vec \tau^\perp_j \ra \!>\! 0$
will favor ferro-octupolar (\fO) order, while $\la \vec \tau^\perp_i \cdot \vec \tau^\perp_j \ra \!<\! 0$ will favor N\'eel octupolar (\nO) order; the 
\fO~and \nO~orders get switched when we consider $K<0$.

Motivated by constructing the simplest model to capture the phenomenology of 
PrTM$_2$Al$_{20}$, we will set $J_1\!<\! 0$ for PrTi$_2$Al$_{20}$ which favors \fQ~order, and $J_1 \! >\!  0$ for PrV$_2$Al$_{20}$ favoring \nQ~order.
In both cases, we fix 
$J_2 \!>\! 0$ and $K \!>\! 0$, and study the phases and their properties as we vary $J_2/|J_1|$ and $K/|J_1|$. At the classical level of the analysis
done here, we note that the model with $J_1 < 0$ maps onto the model with $J_1\!>\! 0$ by changing $\vec \tau \to -\vec \tau$ on one sublattice;
with this understanding, we will mainly focus on fixed $J_1=+1$, but present results which are applicable for both systems.

\begin{figure}[t]
	\centering
	\includegraphics[width=\columnwidth]{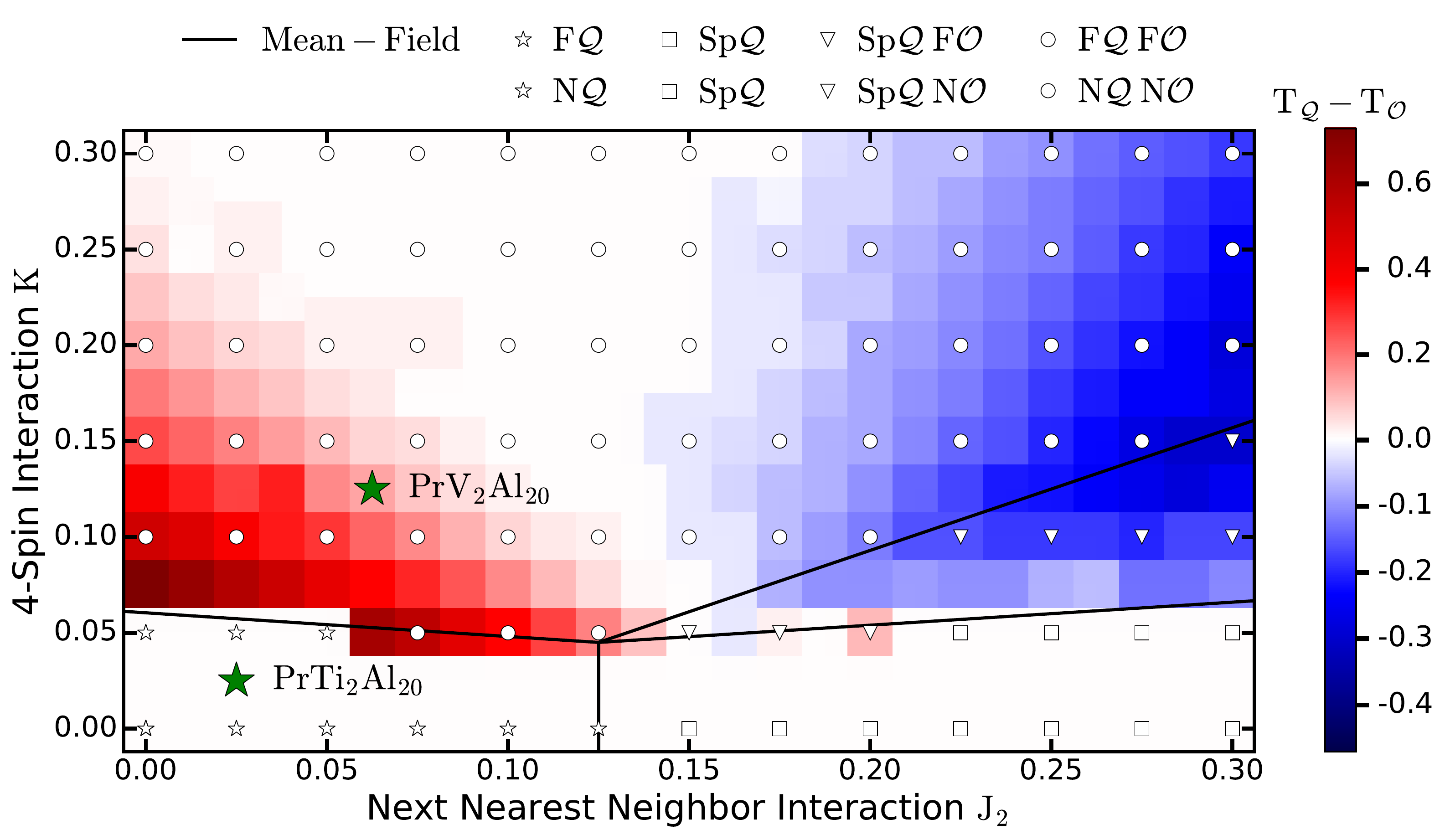}
	\caption{{\bf Ground state phase diagram} for the $J_1$-$J_2$-$K$ model 
	for fixed $J_1=+1$, showing various ordered quadrupolar phases 
	(\nQ=N\'eel quadrupolar, \spQ=spiral quadrupolar)
	as well as coexisting octupolar order (\nO=N\'eel octupolar). For $J_1=-1$, the phase diagram is identical but phases get
	relabelled as \nQ~$\to$ \fQ~(ferroquadrupolar)
	and \nO $\to$ \fO~(ferrooctupolar).
	Solid lines are $T=0$ mean field phase boundaries,
	points are obtained from Monte Carlo (MC) simulations on system sizes $L=8$ ($1024$ spins) showing excellent agreement. 
	Color indicates regions where we find two-stage thermal ordering in MC; the scale shows which broken symmetry 
	(quadrupolar/octupolar) has a higher transition temperature. The
        ``stars" indicate regions where we tentatively place the PrTM$_2$Al$_{20}$ materials 
        (with $J_1<0$ for PrTi$_2$Al$_{20}$ and $J_1 > 0$ for PrV$_2$Al$_{20}$).
	\label{fig:JKmodel-phase-diagram}
		}
	\label{fig:finiteT-phase-diag}
\end{figure}


{\em Ground state phase diagram.---}
For $J_1 > 0$, consider an ansatz 
$\vec \tau^+_{A/B} \!\!=\!\! \sqrt{1\!-\! \eta^2} \exp(i \bq \cdot \br \! \pm \! \frac{\phi}{2})$ for unit length spins on
A/B sublattices, with
$\tau^z_{A/B} \!=\! \pm \eta$. Here $\bq, \phi$ specify a spiral of $\vec\tau^\perp$ which is a generic \spQ~order with
magnitude $\sqrt{1-\eta^2}$. The limit
$\bQ=0$ corresponds to the \nQ~state.
This
coexists with \nO~order of strength $\eta$. Let us define
$F \! \equiv \! \cos \phi \cos \frac{q_x}{4}\cos \frac{q_y}{4}\cos \frac{q_z}{4}
\!-\! \sin\phi \sin \frac{q_x}{4}\sin \frac{q_y}{4}\sin \frac{q_z}{4}$
and $G \! \equiv \! \cos \frac{q_x}{2}\cos \frac{q_y}{2}+ \cos \frac{q_y}{2}\cos \frac{q_z}{2}+\cos \frac{q_z}{2}\cos \frac{q_x}{2}$,
in terms of which we find the energy per site in the classical limit
\bea
\frac{E_{\rm cl}}{N_{\rm site}} &=& -2 (J_1 - 18 K \eta^2)  (1-\eta^2) F(\phi,\bq)
+ 2 J_1 \lambda \eta^2 \nonumber \\
&+& 6 J_2 \lambda \eta^2 + 2 J_2 (1-\eta^2) G(\bq) \,.
\eea
Minimizing this variational energy with respect to $(\bq, \phi, \eta)$, 
we arrive at the $T=0$ phase diagram, with phase boundaries depicted by solid lines in Fig.~\ref{fig:finiteT-phase-diag} for the
choice $\lambda=0$.
Along the line $K=0$, this phase diagram is identical with previous results obtained for Heisenberg spins on the diamond lattice,
where $J_2/J_1 \! > \! 1/8$ drives a N\'eel to incommensurate spiral transition
\cite{bergman2007order,lee2008theory}. Our new results show that $K \neq 0 $ can
induce \nQ/\spQ~phases which coexist with Ising \nO~order; we find qualitatively similar results for generic $\lambda < 1$ 
(see Supplemental Material). For $J_1<0$, the 
\nQ/\nO~phases get replaced by \fQ/\fO~phases, while the spiral is modified by flipping $\vec\tau$
on one sublattice.

\begin{figure}[t]
	\centering
	\includegraphics[width=\linewidth]{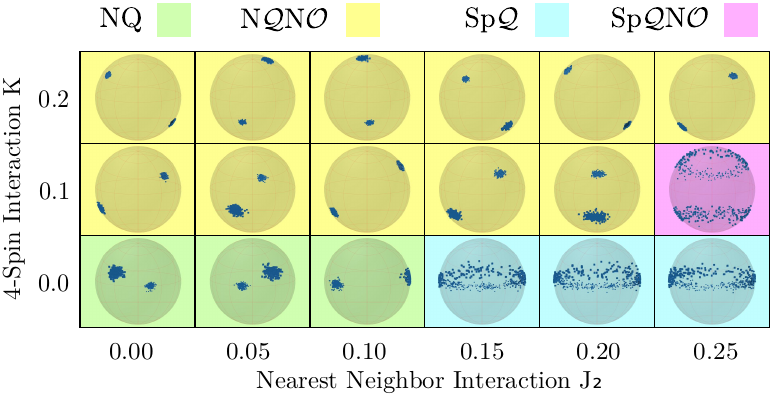}
	\caption{{\bf Common-origin plots} of the spin vectors in configuration snapshots from Monte Carlo simulations
			visualize the  nature of the low-temperature ordering in the ground-state phase diagram 
			of the $J_1$-$J_2$-$K$ model in the $J_2$-$K$ plane.
		}
	\label{fig:mc-phase-diag}
\end{figure}

We have checked the $T=0$ phase diagram in Fig.~\ref{fig:finiteT-phase-diag} using classical Monte Carlo (MC) simulations for system sizes up to
$L\!=\! 8$ (with $2L^3\!=\! 1024$ spins) down to $T/J_1\!=\! 0.001$
at a large number of depicted points.
The distinct ground states are best visualized in common origin plots of the spin vectors of configuration snapshots in the MC simulation as shown in 
Fig.~\ref{fig:mc-phase-diag}. Depending on the $\tau^z$-order of the phase, characteristic $\vec\tau^\perp$-features (such as a ring for the 
spiral phase) are shifted along the z-axis in the common origin plot. The MC simulations clearly confirm our mean 
field ground state phase diagram.


\begin{figure}[t]
	\centering
	\includegraphics[width=\columnwidth]{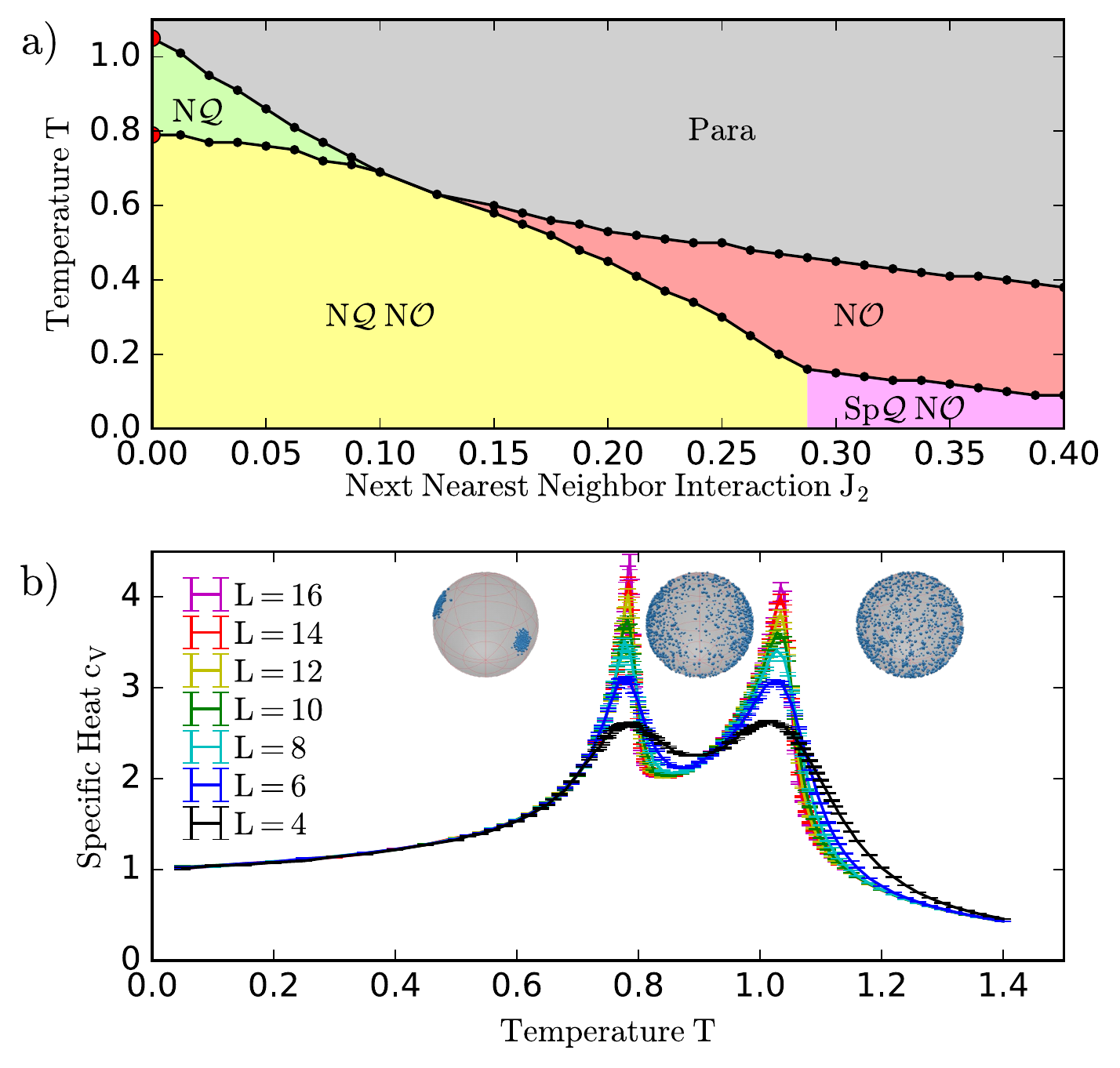}
	\caption{(a) {\bf Finite-temperature phase diagram} as a function of $J_2$ for fixed $J_1=+1,K=0.15$. The phase
	diagram is deduced from specific heat calculations which detects the phase transitions, and from common origin plots which show the nature
	of the phases.
	(b) Illustrative plot of the specific heat versus temperature for $J_1=+1$, $K = 0.15$, for fixed $J_2=0$ (in the \nQ\nO~phase) for various system sizes.
	Also shown are the common origin plots depicting the evolution from paramagnetic to \nQ~to \nQ\nO~order.}
	\label{fig:CV_app2}
\end{figure}

{\em Thermal transitions.---}
In order to explore the phase diagram of this model at nonzero temperatures, we have carried out extensive MC simulations for various system sizes and across a 
broad temperature regime. Fig.~\ref{fig:CV_app2}(a) shows the phase diagram in the $J_2$-$T$ plane at fixed $J_1=1,K=0.15$. We find that both the \nQ\nO~and
the \spQ\nO~phases generically undergo multiple phase transitions enroute to the high temperature paramagnet, with intervening phases which have pure octupolar or
quadrupolar order. We deduce the existence of such transitions via peaks in the specific heat versus temperature, as illustrated in 
Fig.~\ref{fig:CV_app2}(b) for $J_2=0$, which get sharper with increasing system size. The nature of the phases can be deduced from common origin plots of
snapshot MC configurations as shown for the \nQ\nO, \nQ, and paramagnetic phases in Fig.~\ref{fig:CV_app2}(b). Using extensive MC simulations of this sort 
over a wide range of parameters, we have compiled a detailed map of the two phase transitions,
as shown in Fig.~\ref{fig:finiteT-phase-diag} with the color scale indicating regions 
where, upon lowering temperature, quadrupolar
$\vec\tau^\perp$ orders first (red, $T_{\cal Q} - T_{\cal O} > 0$)  or octupolar $\tau^z$ orders first (blue, $T_{\cal Q} - T_{\cal O} < 0$).

{\em Magnetic field effect.---}
We next turn to the impact of an applied magnetic field as a further way to distinguish \fQ~from \afQ~order. We begin by noting that
the quadrupolar and octupolar moments of the Pr$^{3+}$ $\Gamma_3$ doublet do not linearly couple to the magnetic field. The leading term is a quadratic-in-field
coupling to 
the quadrupolar moment originating at second order perturbation theory in $\vec{h} \cdot \vec{J}$. This leads to nonzero matrix elements
in the $\Gamma_3$ doublet with intermediate states arising from excited crystal field levels as 
$H_{h} =\sum_{\alpha}  \frac{ \vec{h} \cdot \vec{J} | \alpha \rangle \langle \alpha |  \vec{h} \cdot \vec{J}} 
{ \Delta(\alpha)} =  \gamma h^2
 \Big( \frac{\sqrt{3}}{2} ( \hat{h}_x^2 - \hat{h}_y^2)  \tau_x 
 + \frac{1}{2} ( 3\hat{h}_z^2 -1) \tau_y  \Big)$ where $\alpha \in {\Gamma_{4},\Gamma_5}$ refers to the 
 two excited triplets above the ground state, $\Delta(\alpha)$ are the corresponding crystal field excitation energies, and
 $\gamma= \Big(- \frac{14}{3 \Delta({\Gamma_4})}  + \frac{2}{\Delta({\Gamma_5})} \Big)$.
The form of the coupling is simply understood on symmetry grounds; since the quadrupolar moments transform like an $e_g$ doublet, 
the magnetic field couples to these moments with the same symmetries. 
Our model Eq.~\eqref{eq:JKmodel} has an $XY$ symmetry, so that magnetic fields along (100) direction or (110) direction act in an identical manner.
However, the quadratic-in-field coupling to the quadrupole moment vanishes for a magnetic field along the (111) direction; 
instead, for this direction, the dominant term is a cubic-in-field coupling
$\sim h^3 (\hat{h}_x \hat{h}_y \hat{h}_z) \tau_z$ to the octupolar moment. 

\begin{figure}[t]
	\centering
	\includegraphics[width=\columnwidth]{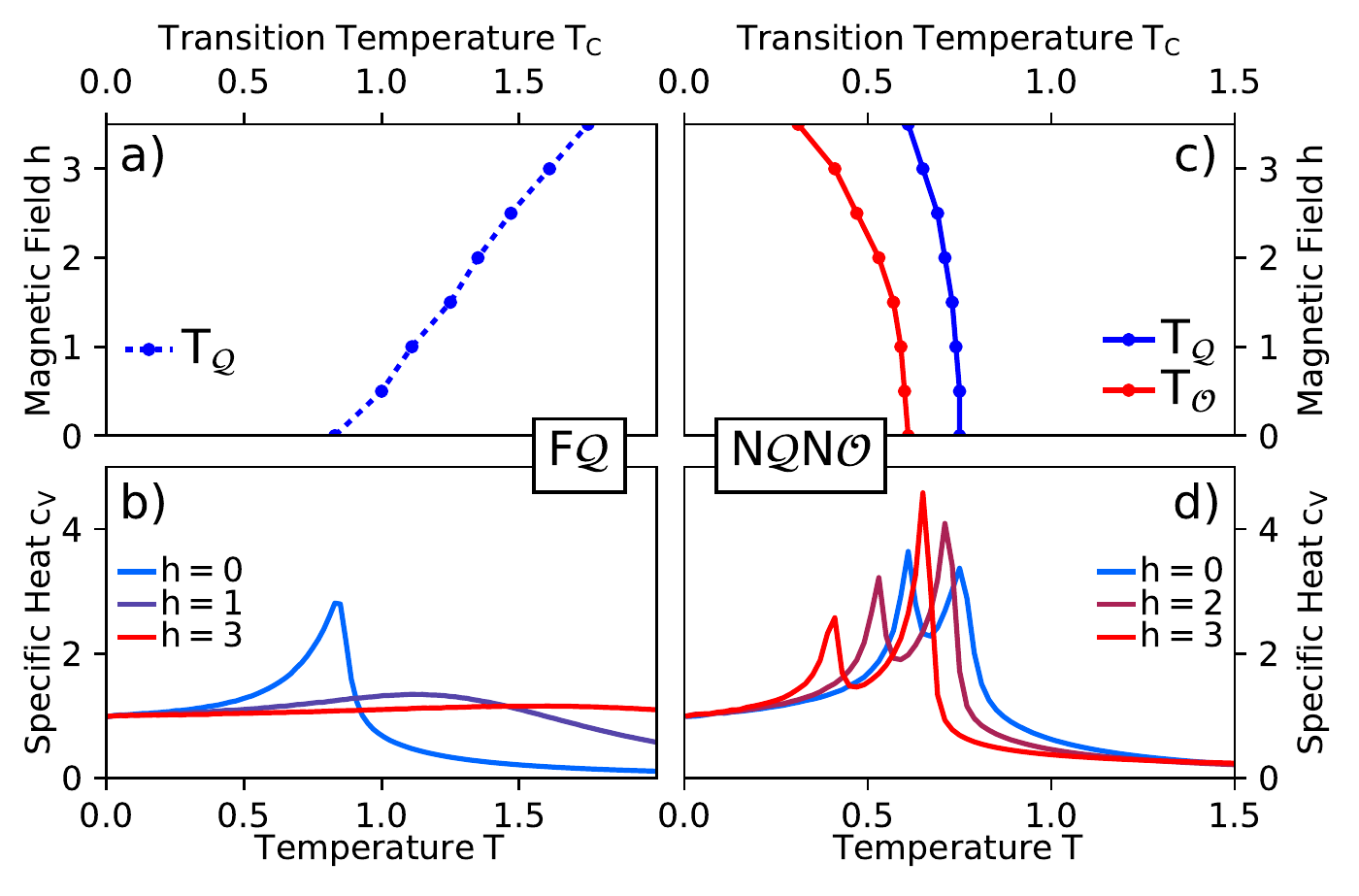}
	\caption{{\bf Response of the \fQ~and \nQ\nO~phases to a [100] magnetic field}. 
			(a) Evolution of the transition temperature 	$T_{\cal Q}$ for the \fQ~state at zero field into a crossover line
			for nonzero field along [100] direction. 
			The crossover temperature in (a) is obtained from specific heat scans as shown in (b), where the sharp peak 
			signaling the transition at zero field becomes a rounded peak for nonzero $h$.  
			(c) Evolution of transition temperatures $T_{\cal Q}$ and $T_{\cal O}$ for the \nQ\nO~state. 
			In this case, the zero field transitions, signaled by the sharp specific heat peaks in (d), survive even for $h\neq 0$, 
			with the field suppressing $T_{\cal O}$ more strongly than $T_{\cal Q}$.}
	\label{fig:mag_field}
\end{figure}

In order to illustrate the effect of the dominant coupling to the quadrupolar order,
Fig.~\ref{fig:mag_field} shows the (100) magnetic field dependence for the \fQ~and \nQ\nO~phases which are presumed to be relevant to
PrTi$_2$Al$_{20}$ and PrV$_2$Al$_{20}$, respectively. In the absence of an applied field, there is a direct
continuous transition from the paramagnet into the \fQ~phase, but the (100) magnetic field converts this into a crossover, 
the crossover temperature increasing with the  field as seen in Figs.~\ref{fig:mag_field}a) and b).
On the other hand, for the \nQ\nO~ phase, both the phase transitions (paramagnet to \nQ~and \nQ~to \nQ\nO)
survive, and the transition temperatures decrease with increasing field. For this model, we find that
the lower temperature transition (\nQ~to \nQ\nO) decreases more rapidly than the higher temperature transition. This can be understood based on
Landau theory which will be discussed in Ref.~\onlinecite{inpreparation} along with a detailed analysis for other field directions.

\noindent
{\em Comparison to experiment.---}
PrTi$_2$Al$_{20}$ exhibits a single phase transition from the paramagnetic phase into a broken symmetry \fQ~phase at $T_c \approx 2$K, as identified
from the fact that the sharp transition becomes a crossover in the presence of a magnetic field \cite{sato2012ferroquadrupolar,onimaru2016exotic}.
As seen in Fig.~\ref{fig:JKmodel-phase-diagram}, the phase diagram with a ferromagnetic $J_1$ and a small $J_2,K > 0$ 
shows a (white) region with a single transition from the paramagnet into the \fQ~phase, which becomes a crossover in a nonzero (100) field as shown
above. We thus place the parameters for the pseudospin-1/2 model for PrTi$_2$Al$_{20}$ in this region.
Contrary to a single phase transition seen in PrTi$_2$Al$_{20}$, 
there exist two phase transitions in the case of PrV$_2$Al$_{20}$ \cite{sakai2011kondo,tsujimoto2015anomalous}.
In addition to $T_1 \approx 0.8$K  for the transition to \nQ~ ordering, it  has been observed that there is another phase transition slightly lower at $T_2 \approx 0.7 $K. It is possible that such two phase transitions originate from 
the ordering of quadrupolar moments (\nQ) and of octupolar moments (\nO) respectively. Again, our model with $J_1 > 0$ and with somewhat larger $J_2, K > 0$
does show a double transition, from paramagnet to \nQ, followed by a lower transition from \nQ~to \nQ\nO. 
There are extended parameter regimes seen in 
Fig.~\ref{fig:JKmodel-phase-diagram} with $J_1 > 0$
(light pink) where such closely spaced double transitions appear; thus we tentatively place PrV$_2$Al$_{20}$ in this regime of the phase diagram.


{\em Discussion.---}
In this Letter, we have argued
that multi-spin interactions should be generically important in Kondo materials near the ordered-to-hybridized FL transition. We have shown that this can lead to coexistence of
quadrupolar and octupolar orders in the Pr(TM)$_2$Al$_{20}$ systems.
If we assume that PrV$_2$Al$_{20}$ has a stronger Kondo-hybridization compared to PrTi$_2$Al$_{20}$, the two-stage thermal transitions seen in PrV$_2$Al$_{20}$ and a single transition in PrTi$_2$Al$_{20}$ would naturally be explained by relative importance of the multi-spin interactions in PrV$_2$Al$_{20}$
or the proximity to the ordered-to-hybridized FL transition. 
Further experiments and theory are needed to explore the dependence of the ordering temperatures on magnetic fields along various directions, which
would further clarify the nature of the broken symmetries and the full phase diagram. Experiments to detect the octupolar order would also be invaluable. 
In this context, we note that $\mu$SR measurements
to look for time-reversal breaking might be challenging since the electric field produced by the muon would break the non-Kramers degeneracy of the
$\Gamma_3$ doublet for nearby Pr$^{3+}$ ions. Nuclear magnetic resonance experiments might provide a complementary tool to detect the octupolar
order.
Finally, the presence of both quadrupolar and octupolar order
may impact the non-Fermi liquid behavior near the putative ordered-to-hybridized FL
quantum critical point. 
The pronounced non-Fermi liquid behavior seen above the multipolar ordering temperature in PrV$_2$Al$_{20}$ may be 
the signatures of such a quantum critical point. 
Future work could explore the coupling between such unusual
order parameters and conduction electrons, which can
lead to novel quantum critical behavior.


\begin{acknowledgments}

\noindent
{\em Acknowledgments.---}
The Cologne group acknowledges partial funding from the DFG within CRC 1238 (project C02).
The numerical simulations were performed on the CHEOPS cluster at RRZK Cologne.
J. A. thanks the Bonn-Cologne Graduate School of Physics and Astronomy (BCGS) for support.
S.B.L. is supported by the KAIST startup and National Research Foundation Grant (NRF-2017R1A2B4008097). 
A.P. and Y.B.K. are supported by the NSERC of Canada and the Canadian Institute for Advanced Research.
S.B.L., A.P., S.T., and Y.B.K. acknowledge hospitality of the ``Intertwined orders" program at the 
Kavli Institute for Theoretical Physics, supported in part by the National Science Foundation 
under Grant No. NSF PHY-1125915. Y.B.K. thanks the hospitality at the Aspen Center for Physics, 
supported in part by NSF grant PHY-1607611. 
\end{acknowledgments}


\bibliography{multipolar}

\begin{thebibliography}{51}%
\makeatletter
\providecommand \@ifxundefined [1]{%
 \@ifx{#1\undefined}
}%
\providecommand \@ifnum [1]{%
 \ifnum #1\expandafter \@firstoftwo
 \else \expandafter \@secondoftwo
 \fi
}%
\providecommand \@ifx [1]{%
 \ifx #1\expandafter \@firstoftwo
 \else \expandafter \@secondoftwo
 \fi
}%
\providecommand \natexlab [1]{#1}%
\providecommand \enquote  [1]{``#1''}%
\providecommand \bibnamefont  [1]{#1}%
\providecommand \bibfnamefont [1]{#1}%
\providecommand \citenamefont [1]{#1}%
\providecommand \href@noop [0]{\@secondoftwo}%
\providecommand \href [0]{\begingroup \@sanitize@url \@href}%
\providecommand \@href[1]{\@@startlink{#1}\@@href}%
\providecommand \@@href[1]{\endgroup#1\@@endlink}%
\providecommand \@sanitize@url [0]{\catcode `\\12\catcode `\$12\catcode
  `\&12\catcode `\#12\catcode `\^12\catcode `\_12\catcode `\%12\relax}%
\providecommand \@@startlink[1]{}%
\providecommand \@@endlink[0]{}%
\providecommand \url  [0]{\begingroup\@sanitize@url \@url }%
\providecommand \@url [1]{\endgroup\@href {#1}{\urlprefix }}%
\providecommand \urlprefix  [0]{URL }%
\providecommand \Eprint [0]{\href }%
\providecommand \doibase [0]{http://dx.doi.org/}%
\providecommand \selectlanguage [0]{\@gobble}%
\providecommand \bibinfo  [0]{\@secondoftwo}%
\providecommand \bibfield  [0]{\@secondoftwo}%
\providecommand \translation [1]{[#1]}%
\providecommand \BibitemOpen [0]{}%
\providecommand \bibitemStop [0]{}%
\providecommand \bibitemNoStop [0]{.\EOS\space}%
\providecommand \EOS [0]{\spacefactor3000\relax}%
\providecommand \BibitemShut  [1]{\csname bibitem#1\endcsname}%
\let\auto@bib@innerbib\@empty
\bibitem [{\citenamefont {Doniach}(1977)}]{doniach1977kondo}%
  \BibitemOpen
  \bibfield  {author} {\bibinfo {author} {\bibfnamefont {S.}~\bibnamefont
  {Doniach}},\ }\bibfield  {title} {\enquote {\bibinfo {title} {{The Kondo
  lattice and weak antiferromagnetism}},}\ }\href@noop {} {\bibfield  {journal}
  {\bibinfo  {journal} {Physica B+C}\ }\textbf {\bibinfo {volume} {91}},\
  \bibinfo {pages} {231} (\bibinfo {year} {1977})}\BibitemShut {NoStop}%
\bibitem [{\citenamefont {Ruderman}\ and\ \citenamefont
  {Kittel}(1954)}]{ruderman1954indirect}%
  \BibitemOpen
  \bibfield  {author} {\bibinfo {author} {\bibfnamefont {Melvin~A.}\
  \bibnamefont {Ruderman}}\ and\ \bibinfo {author} {\bibfnamefont {Charles}\
  \bibnamefont {Kittel}},\ }\bibfield  {title} {\enquote {\bibinfo {title}
  {Indirect exchange coupling of nuclear magnetic moments by conduction
  electrons},}\ }\href@noop {} {\bibfield  {journal} {\bibinfo  {journal}
  {Phys. Rev.}\ }\textbf {\bibinfo {volume} {96}},\ \bibinfo {pages} {99}
  (\bibinfo {year} {1954})}\BibitemShut {NoStop}%
\bibitem [{\citenamefont {Kasuya}(1956)}]{kasuya1956theory}%
  \BibitemOpen
  \bibfield  {author} {\bibinfo {author} {\bibfnamefont {Tadao}\ \bibnamefont
  {Kasuya}},\ }\bibfield  {title} {\enquote {\bibinfo {title} {{A theory of
  metallic ferro-and antiferromagnetism on Zener's model}},}\ }\href@noop {}
  {\bibfield  {journal} {\bibinfo  {journal} {Progress of Theoretical Physics}\
  }\textbf {\bibinfo {volume} {16}},\ \bibinfo {pages} {45} (\bibinfo {year}
  {1956})}\BibitemShut {NoStop}%
\bibitem [{\citenamefont {Yosida}(1957)}]{yosida1957magnetic}%
  \BibitemOpen
  \bibfield  {author} {\bibinfo {author} {\bibfnamefont {Kei}\ \bibnamefont
  {Yosida}},\ }\bibfield  {title} {\enquote {\bibinfo {title} {{Magnetic
  properties of Cu-Mn alloys}},}\ }\href@noop {} {\bibfield  {journal}
  {\bibinfo  {journal} {Phys. Rev.}\ }\textbf {\bibinfo {volume} {106}},\
  \bibinfo {pages} {893} (\bibinfo {year} {1957})}\BibitemShut {NoStop}%
\bibitem [{\citenamefont {Stewart}(1984)}]{stewart1984heavy}%
  \BibitemOpen
  \bibfield  {author} {\bibinfo {author} {\bibfnamefont {See~G.R.}\
  \bibnamefont {Stewart}},\ }\bibfield  {title} {\enquote {\bibinfo {title}
  {Heavy-fermion systems},}\ }\href@noop {} {\bibfield  {journal} {\bibinfo
  {journal} {Rev. Mod. Phys.}\ }\textbf {\bibinfo {volume} {56}},\ \bibinfo
  {pages} {755} (\bibinfo {year} {1984})}\BibitemShut {NoStop}%
\bibitem [{\citenamefont {Fisk}\ \emph {et~al.}(1995)\citenamefont {Fisk},
  \citenamefont {Sarrao}, \citenamefont {Smith},\ and\ \citenamefont
  {Thompson}}]{fisk1995physics}%
  \BibitemOpen
  \bibfield  {author} {\bibinfo {author} {\bibfnamefont {Z.}~\bibnamefont
  {Fisk}}, \bibinfo {author} {\bibfnamefont {J.L.}\ \bibnamefont {Sarrao}},
  \bibinfo {author} {\bibfnamefont {J.L.}\ \bibnamefont {Smith}}, \ and\
  \bibinfo {author} {\bibfnamefont {J.D.}\ \bibnamefont {Thompson}},\
  }\bibfield  {title} {\enquote {\bibinfo {title} {The physics and chemistry of
  heavy fermions},}\ }\href@noop {} {\bibfield  {journal} {\bibinfo  {journal}
  {Proceedings of the National Academy of Sciences}\ }\textbf {\bibinfo
  {volume} {92}},\ \bibinfo {pages} {6663} (\bibinfo {year}
  {1995})}\BibitemShut {NoStop}%
\bibitem [{\citenamefont {Coleman}\ \emph {et~al.}(2001)\citenamefont
  {Coleman}, \citenamefont {P{\'e}pin}, \citenamefont {Si},\ and\ \citenamefont
  {Ramazashvili}}]{coleman2001fermi}%
  \BibitemOpen
  \bibfield  {author} {\bibinfo {author} {\bibfnamefont {P.}~\bibnamefont
  {Coleman}}, \bibinfo {author} {\bibfnamefont {C.}~\bibnamefont {P{\'e}pin}},
  \bibinfo {author} {\bibfnamefont {Qimiao}\ \bibnamefont {Si}}, \ and\
  \bibinfo {author} {\bibfnamefont {Revaz}\ \bibnamefont {Ramazashvili}},\
  }\bibfield  {title} {\enquote {\bibinfo {title} {How do fermi liquids get
  heavy and die?}}\ }\href@noop {} {\bibfield  {journal} {\bibinfo  {journal}
  {Journal of Physics: Condensed Matter}\ }\textbf {\bibinfo {volume} {13}},\
  \bibinfo {pages} {R723} (\bibinfo {year} {2001})}\BibitemShut {NoStop}%
\bibitem [{\citenamefont {Gegenwart}\ \emph {et~al.}(2008)\citenamefont
  {Gegenwart}, \citenamefont {Si},\ and\ \citenamefont
  {Steglich}}]{gegenwart2008quantum}%
  \BibitemOpen
  \bibfield  {author} {\bibinfo {author} {\bibfnamefont {Philipp}\ \bibnamefont
  {Gegenwart}}, \bibinfo {author} {\bibfnamefont {Qimiao}\ \bibnamefont {Si}},
  \ and\ \bibinfo {author} {\bibfnamefont {Frank}\ \bibnamefont {Steglich}},\
  }\bibfield  {title} {\enquote {\bibinfo {title} {Quantum criticality in
  heavy-fermion metals},}\ }\href@noop {} {\bibfield  {journal} {\bibinfo
  {journal} {Nature Physics}\ }\textbf {\bibinfo {volume} {4}},\ \bibinfo
  {pages} {186} (\bibinfo {year} {2008})}\BibitemShut {NoStop}%
\bibitem [{\citenamefont {Si}\ and\ \citenamefont
  {Steglich}(2010)}]{si2010heavy}%
  \BibitemOpen
  \bibfield  {author} {\bibinfo {author} {\bibfnamefont {Qimiao}\ \bibnamefont
  {Si}}\ and\ \bibinfo {author} {\bibfnamefont {Frank}\ \bibnamefont
  {Steglich}},\ }\bibfield  {title} {\enquote {\bibinfo {title} {Heavy fermions
  and quantum phase transitions},}\ }\href@noop {} {\bibfield  {journal}
  {\bibinfo  {journal} {Science}\ }\textbf {\bibinfo {volume} {329}},\ \bibinfo
  {pages} {1161} (\bibinfo {year} {2010})}\BibitemShut {NoStop}%
\bibitem [{\citenamefont {Morin}\ \emph {et~al.}(1982)\citenamefont {Morin},
  \citenamefont {Schmitt},\ and\ \citenamefont
  {De~Lacheisserie}}]{morin1982magnetic}%
  \BibitemOpen
  \bibfield  {author} {\bibinfo {author} {\bibfnamefont {P.}~\bibnamefont
  {Morin}}, \bibinfo {author} {\bibfnamefont {D.}~\bibnamefont {Schmitt}}, \
  and\ \bibinfo {author} {\bibfnamefont {E.~Du~Tremolet}\ \bibnamefont
  {De~Lacheisserie}},\ }\bibfield  {title} {\enquote {\bibinfo {title}
  {{Magnetic and quadrupolar properties of PrPb$_3$}},}\ }\href@noop {}
  {\bibfield  {journal} {\bibinfo  {journal} {Journal of Magnetism and Magnetic
  Materials}\ }\textbf {\bibinfo {volume} {30}},\ \bibinfo {pages} {257}
  (\bibinfo {year} {1982})}\BibitemShut {NoStop}%
\bibitem [{\citenamefont {Cox}(1987)}]{cox1987quadrupolar}%
  \BibitemOpen
  \bibfield  {author} {\bibinfo {author} {\bibfnamefont {D.L.}\ \bibnamefont
  {Cox}},\ }\bibfield  {title} {\enquote {\bibinfo {title} {{Quadrupolar Kondo
  effect in uranium heavy-electron materials?}}}\ }\href@noop {} {\bibfield
  {journal} {\bibinfo  {journal} {Phys. Rev. Lett.}\ }\textbf {\bibinfo
  {volume} {59}},\ \bibinfo {pages} {1240} (\bibinfo {year}
  {1987})}\BibitemShut {NoStop}%
\bibitem [{\citenamefont {Cox}\ and\ \citenamefont
  {Zawadowski}(1999)}]{cox1999exotic}%
  \BibitemOpen
  \bibfield  {author} {\bibinfo {author} {\bibfnamefont {D.L.}\ \bibnamefont
  {Cox}}\ and\ \bibinfo {author} {\bibfnamefont {Alfred}\ \bibnamefont
  {Zawadowski}},\ }\href@noop {} {\emph {\bibinfo {title} {{Exotic Kondo
  Effects in Metals: Magnetic Ions in a Crystalline Electric Field and
  Tunelling Centres}}}}\ (\bibinfo  {publisher} {CRC Press},\ \bibinfo {year}
  {1999})\BibitemShut {NoStop}%
\bibitem [{\citenamefont {Kitagawa}\ \emph {et~al.}(1996)\citenamefont
  {Kitagawa}, \citenamefont {Takeda},\ and\ \citenamefont
  {Ishikawa}}]{kitagawa1996possible}%
  \BibitemOpen
  \bibfield  {author} {\bibinfo {author} {\bibfnamefont {Jiro}\ \bibnamefont
  {Kitagawa}}, \bibinfo {author} {\bibfnamefont {Naoya}\ \bibnamefont
  {Takeda}}, \ and\ \bibinfo {author} {\bibfnamefont {Masayasu}\ \bibnamefont
  {Ishikawa}},\ }\bibfield  {title} {\enquote {\bibinfo {title} {{Possible
  quadrupolar ordering in a Kondo-lattice compound Ce$_3$Pd$_{20}$Ge$_6$}},}\
  }\href@noop {} {\bibfield  {journal} {\bibinfo  {journal} {Phys. Rev. B}\
  }\textbf {\bibinfo {volume} {53}},\ \bibinfo {pages} {5101} (\bibinfo {year}
  {1996})}\BibitemShut {NoStop}%
\bibitem [{\citenamefont {Caciuffo}\ \emph {et~al.}(2003)\citenamefont
  {Caciuffo}, \citenamefont {Paix{\~a}o}, \citenamefont {Detlefs},
  \citenamefont {Longfield}, \citenamefont {Santini}, \citenamefont
  {Bernhoeft}, \citenamefont {Rebizant},\ and\ \citenamefont
  {Lander}}]{caciuffo2003multipolar}%
  \BibitemOpen
  \bibfield  {author} {\bibinfo {author} {\bibfnamefont {R.}~\bibnamefont
  {Caciuffo}}, \bibinfo {author} {\bibfnamefont {J.A.}\ \bibnamefont
  {Paix{\~a}o}}, \bibinfo {author} {\bibfnamefont {C.}~\bibnamefont {Detlefs}},
  \bibinfo {author} {\bibfnamefont {M.J.}\ \bibnamefont {Longfield}}, \bibinfo
  {author} {\bibfnamefont {P.}~\bibnamefont {Santini}}, \bibinfo {author}
  {\bibfnamefont {N.}~\bibnamefont {Bernhoeft}}, \bibinfo {author}
  {\bibfnamefont {J.}~\bibnamefont {Rebizant}}, \ and\ \bibinfo {author}
  {\bibfnamefont {G.H.}\ \bibnamefont {Lander}},\ }\bibfield  {title} {\enquote
  {\bibinfo {title} {{Multipolar ordering in NpO$_2$ below 25 K}},}\
  }\href@noop {} {\bibfield  {journal} {\bibinfo  {journal} {Journal of
  Physics: Condensed Matter}\ }\textbf {\bibinfo {volume} {15}},\ \bibinfo
  {pages} {S2287} (\bibinfo {year} {2003})}\BibitemShut {NoStop}%
\bibitem [{\citenamefont {Suzuki}\ \emph {et~al.}(2005)\citenamefont {Suzuki},
  \citenamefont {S.~Suzuki}, \citenamefont {Kitazawa}, \citenamefont {Kido},
  \citenamefont {Ueno}, \citenamefont {Yamaguchi}, \citenamefont {Nemoto},\
  and\ \citenamefont {Goto}}]{suzuki2005quadrupolar}%
  \BibitemOpen
  \bibfield  {author} {\bibinfo {author} {\bibfnamefont {Osamu}\ \bibnamefont
  {Suzuki}}, \bibinfo {author} {\bibfnamefont {Hiroyuki}\ \bibnamefont
  {S.~Suzuki}}, \bibinfo {author} {\bibfnamefont {Hideaki}\ \bibnamefont
  {Kitazawa}}, \bibinfo {author} {\bibfnamefont {Giyuu}\ \bibnamefont {Kido}},
  \bibinfo {author} {\bibfnamefont {Takafumi}\ \bibnamefont {Ueno}}, \bibinfo
  {author} {\bibfnamefont {Takashi}\ \bibnamefont {Yamaguchi}}, \bibinfo
  {author} {\bibfnamefont {Yuichi}\ \bibnamefont {Nemoto}}, \ and\ \bibinfo
  {author} {\bibfnamefont {Terutaka}\ \bibnamefont {Goto}},\ }\bibfield
  {title} {\enquote {\bibinfo {title} {{Quadrupolar Kondo effect in non-Kramers
  doublet system PrInAg$_2$}},}\ }\href@noop {} {\bibfield  {journal} {\bibinfo
   {journal} {J. Phys. Soc. Jpn.}\ }\textbf {\bibinfo {volume} {75}},\ \bibinfo
  {pages} {013704} (\bibinfo {year} {2005})}\BibitemShut {NoStop}%
\bibitem [{\citenamefont {Kuramoto}\ \emph {et~al.}(2009)\citenamefont
  {Kuramoto}, \citenamefont {Kusunose},\ and\ \citenamefont
  {Kiss}}]{kuramoto2009multipole}%
  \BibitemOpen
  \bibfield  {author} {\bibinfo {author} {\bibfnamefont {Yoshio}\ \bibnamefont
  {Kuramoto}}, \bibinfo {author} {\bibfnamefont {Hiroaki}\ \bibnamefont
  {Kusunose}}, \ and\ \bibinfo {author} {\bibfnamefont {Annamaria}\
  \bibnamefont {Kiss}},\ }\bibfield  {title} {\enquote {\bibinfo {title}
  {Multipole orders and fluctuations in strongly correlated electron
  systems},}\ }\href@noop {} {\bibfield  {journal} {\bibinfo  {journal} {J.
  Phys. Soc. Jpn.}\ }\textbf {\bibinfo {volume} {78}},\ \bibinfo {pages}
  {072001} (\bibinfo {year} {2009})}\BibitemShut {NoStop}%
\bibitem [{\citenamefont {Lee}\ \emph {et~al.}(2015)\citenamefont {Lee},
  \citenamefont {Paramekanti},\ and\ \citenamefont {Kim}}]{lee2015optical}%
  \BibitemOpen
  \bibfield  {author} {\bibinfo {author} {\bibfnamefont {SungBin}\ \bibnamefont
  {Lee}}, \bibinfo {author} {\bibfnamefont {Arun}\ \bibnamefont {Paramekanti}},
  \ and\ \bibinfo {author} {\bibfnamefont {Yong~Baek}\ \bibnamefont {Kim}},\
  }\bibfield  {title} {\enquote {\bibinfo {title} {{Optical gyrotropy in
  quadrupolar Kondo systems}},}\ }\href@noop {} {\bibfield  {journal} {\bibinfo
   {journal} {Phys. Rev. B}\ }\textbf {\bibinfo {volume} {91}},\ \bibinfo
  {pages} {041104} (\bibinfo {year} {2015})}\BibitemShut {NoStop}%
\bibitem [{\citenamefont {Sakai}\ and\ \citenamefont
  {Nakatsuji}(2011)}]{sakai2011kondo}%
  \BibitemOpen
  \bibfield  {author} {\bibinfo {author} {\bibfnamefont {Akito}\ \bibnamefont
  {Sakai}}\ and\ \bibinfo {author} {\bibfnamefont {Satoru}\ \bibnamefont
  {Nakatsuji}},\ }\bibfield  {title} {\enquote {\bibinfo {title} {{Kondo
  Effects and Multipolar Order in the Cubic PrTr$_2$Al$_{20}$ (Tr=Ti, V)}},}\
  }\href@noop {} {\bibfield  {journal} {\bibinfo  {journal} {J. Phys. Soc.
  Jpn.}\ }\textbf {\bibinfo {volume} {80}},\ \bibinfo {pages} {063701}
  (\bibinfo {year} {2011})}\BibitemShut {NoStop}%
\bibitem [{\citenamefont {Koseki}\ \emph {et~al.}(2011)\citenamefont {Koseki},
  \citenamefont {Nakanishi}, \citenamefont {Deto}, \citenamefont {Koseki},
  \citenamefont {Kashiwazaki}, \citenamefont {Shichinomiya}, \citenamefont
  {Nakamura}, \citenamefont {Yoshizawa}, \citenamefont {Sakai},\ and\
  \citenamefont {Nakatsuji}}]{koseki2011ultrasonic}%
  \BibitemOpen
  \bibfield  {author} {\bibinfo {author} {\bibfnamefont {Minoru}\ \bibnamefont
  {Koseki}}, \bibinfo {author} {\bibfnamefont {Yoshiki}\ \bibnamefont
  {Nakanishi}}, \bibinfo {author} {\bibfnamefont {Kazuhisa}\ \bibnamefont
  {Deto}}, \bibinfo {author} {\bibfnamefont {Gen}\ \bibnamefont {Koseki}},
  \bibinfo {author} {\bibfnamefont {Reiko}\ \bibnamefont {Kashiwazaki}},
  \bibinfo {author} {\bibfnamefont {Fumitaka}\ \bibnamefont {Shichinomiya}},
  \bibinfo {author} {\bibfnamefont {Mitsuteru}\ \bibnamefont {Nakamura}},
  \bibinfo {author} {\bibfnamefont {Masahito}\ \bibnamefont {Yoshizawa}},
  \bibinfo {author} {\bibfnamefont {Akihito}\ \bibnamefont {Sakai}}, \ and\
  \bibinfo {author} {\bibfnamefont {Satoru}\ \bibnamefont {Nakatsuji}},\
  }\bibfield  {title} {\enquote {\bibinfo {title} {{Ultrasonic investigation on
  a cage structure compound PrTi$_2$Al$_{20}$}},}\ }\href@noop {} {\bibfield
  {journal} {\bibinfo  {journal} {J. Phys. Soc. Jpn.}\ }\textbf {\bibinfo
  {volume} {80}},\ \bibinfo {pages} {SA049} (\bibinfo {year}
  {2011})}\BibitemShut {NoStop}%
\bibitem [{\citenamefont {Sakai}\ and\ \citenamefont
  {Nakatsuji}(2012)}]{sakai2012thermal}%
  \BibitemOpen
  \bibfield  {author} {\bibinfo {author} {\bibfnamefont {Akito}\ \bibnamefont
  {Sakai}}\ and\ \bibinfo {author} {\bibfnamefont {Satoru}\ \bibnamefont
  {Nakatsuji}},\ }\bibfield  {title} {\enquote {\bibinfo {title} {{Thermal
  properties of the nonmagnetic cubic $\Gamma$3 Kondo lattice systems
  PrTr$_2$Al$_{20}$ (Tr=Ti, V)}},}\ }in\ \href@noop {} {\emph {\bibinfo
  {booktitle} {Journal of Physics: Conference Series}}},\ Vol.\ \bibinfo
  {volume} {391}\ (\bibinfo {organization} {IOP Publishing},\ \bibinfo {year}
  {2012})\ p.\ \bibinfo {pages} {012058}\BibitemShut {NoStop}%
\bibitem [{\citenamefont {Sato}\ \emph {et~al.}(2012)\citenamefont {Sato},
  \citenamefont {Ibuka}, \citenamefont {Nambu}, \citenamefont {Yamazaki},
  \citenamefont {Hong}, \citenamefont {Sakai},\ and\ \citenamefont
  {Nakatsuji}}]{sato2012ferroquadrupolar}%
  \BibitemOpen
  \bibfield  {author} {\bibinfo {author} {\bibfnamefont {Taku~J.}\ \bibnamefont
  {Sato}}, \bibinfo {author} {\bibfnamefont {Soshi}\ \bibnamefont {Ibuka}},
  \bibinfo {author} {\bibfnamefont {Yusuke}\ \bibnamefont {Nambu}}, \bibinfo
  {author} {\bibfnamefont {Teruo}\ \bibnamefont {Yamazaki}}, \bibinfo {author}
  {\bibfnamefont {Tao}\ \bibnamefont {Hong}}, \bibinfo {author} {\bibfnamefont
  {Akito}\ \bibnamefont {Sakai}}, \ and\ \bibinfo {author} {\bibfnamefont
  {Satoru}\ \bibnamefont {Nakatsuji}},\ }\bibfield  {title} {\enquote {\bibinfo
  {title} {{Ferroquadrupolar ordering in PrTi$_2$Al$_{20}$}},}\ }\href@noop {}
  {\bibfield  {journal} {\bibinfo  {journal} {Phys. Rev. B}\ }\textbf {\bibinfo
  {volume} {86}},\ \bibinfo {pages} {184419} (\bibinfo {year}
  {2012})}\BibitemShut {NoStop}%
\bibitem [{\citenamefont {Onimaru}\ and\ \citenamefont
  {Kusunose}(2016)}]{onimaru2016exotic}%
  \BibitemOpen
  \bibfield  {author} {\bibinfo {author} {\bibfnamefont {Takahiro}\
  \bibnamefont {Onimaru}}\ and\ \bibinfo {author} {\bibfnamefont {Hiroaki}\
  \bibnamefont {Kusunose}},\ }\bibfield  {title} {\enquote {\bibinfo {title}
  {{Exotic Quadrupolar Phenomena in Non-Kramers Doublet Systems? The Cases of
  PrT$_2$Zn$_{20}$ (T=Ir, Rh) and PrT$_2$Al$_{20}$ (T=V, Ti)?}}}\ }\href@noop
  {} {\bibfield  {journal} {\bibinfo  {journal} {J. Phys. Soc. Jpn.}\ }\textbf
  {\bibinfo {volume} {85}},\ \bibinfo {pages} {082002} (\bibinfo {year}
  {2016})}\BibitemShut {NoStop}%
\bibitem [{\citenamefont {Onimaru}\ \emph {et~al.}(2011)\citenamefont
  {Onimaru}, \citenamefont {Matsumoto}, \citenamefont {Inoue}, \citenamefont
  {Umeo}, \citenamefont {Sakakibara}, \citenamefont {Karaki}, \citenamefont
  {Kubota},\ and\ \citenamefont
  {Takabatake}}]{onimaru2011antiferroquadrupolar}%
  \BibitemOpen
  \bibfield  {author} {\bibinfo {author} {\bibfnamefont {T.}~\bibnamefont
  {Onimaru}}, \bibinfo {author} {\bibfnamefont {K.T.}\ \bibnamefont
  {Matsumoto}}, \bibinfo {author} {\bibfnamefont {Y.F.}\ \bibnamefont {Inoue}},
  \bibinfo {author} {\bibfnamefont {K.}~\bibnamefont {Umeo}}, \bibinfo {author}
  {\bibfnamefont {T.}~\bibnamefont {Sakakibara}}, \bibinfo {author}
  {\bibfnamefont {Y.}~\bibnamefont {Karaki}}, \bibinfo {author} {\bibfnamefont
  {M.}~\bibnamefont {Kubota}}, \ and\ \bibinfo {author} {\bibfnamefont
  {T.}~\bibnamefont {Takabatake}},\ }\bibfield  {title} {\enquote {\bibinfo
  {title} {{Antiferroquadrupolar ordering in a Pr-based superconductor
  PrIr$_2$Zn$_20$}},}\ }\href@noop {} {\bibfield  {journal} {\bibinfo
  {journal} {Phys. Rev. Lett.}\ }\textbf {\bibinfo {volume} {106}},\ \bibinfo
  {pages} {177001} (\bibinfo {year} {2011})}\BibitemShut {NoStop}%
\bibitem [{\citenamefont {Shimura}\ \emph {et~al.}(2013)\citenamefont
  {Shimura}, \citenamefont {Ohta}, \citenamefont {Sakakibara}, \citenamefont
  {Sakai},\ and\ \citenamefont {Nakatsuji}}]{shimura2013evidence}%
  \BibitemOpen
  \bibfield  {author} {\bibinfo {author} {\bibfnamefont {Yasuyuki}\
  \bibnamefont {Shimura}}, \bibinfo {author} {\bibfnamefont {Yasuo}\
  \bibnamefont {Ohta}}, \bibinfo {author} {\bibfnamefont {Toshiro}\
  \bibnamefont {Sakakibara}}, \bibinfo {author} {\bibfnamefont {Akito}\
  \bibnamefont {Sakai}}, \ and\ \bibinfo {author} {\bibfnamefont {Satoru}\
  \bibnamefont {Nakatsuji}},\ }\bibfield  {title} {\enquote {\bibinfo {title}
  {{Evidence of a High-Field Phase in PrV$_2$Al${20}$ in a [100] Magnetic
  Field}},}\ }\href@noop {} {\bibfield  {journal} {\bibinfo  {journal} {J.
  Phys. Soc. Jpn.}\ }\textbf {\bibinfo {volume} {82}},\ \bibinfo {pages}
  {043705} (\bibinfo {year} {2013})}\BibitemShut {NoStop}%
\bibitem [{\citenamefont {Onimaru}\ \emph {et~al.}(2012)\citenamefont
  {Onimaru}, \citenamefont {Nagasawa}, \citenamefont {Matsumoto}, \citenamefont
  {Wakiya}, \citenamefont {Umeo}, \citenamefont {Kittaka}, \citenamefont
  {Sakakibara}, \citenamefont {Matsushita},\ and\ \citenamefont
  {Takabatake}}]{onimaru2012simultaneous}%
  \BibitemOpen
  \bibfield  {author} {\bibinfo {author} {\bibfnamefont {T.}~\bibnamefont
  {Onimaru}}, \bibinfo {author} {\bibfnamefont {N.}~\bibnamefont {Nagasawa}},
  \bibinfo {author} {\bibfnamefont {K.T.}\ \bibnamefont {Matsumoto}}, \bibinfo
  {author} {\bibfnamefont {K.}~\bibnamefont {Wakiya}}, \bibinfo {author}
  {\bibfnamefont {K.}~\bibnamefont {Umeo}}, \bibinfo {author} {\bibfnamefont
  {S.}~\bibnamefont {Kittaka}}, \bibinfo {author} {\bibfnamefont
  {T.}~\bibnamefont {Sakakibara}}, \bibinfo {author} {\bibfnamefont
  {Y.}~\bibnamefont {Matsushita}}, \ and\ \bibinfo {author} {\bibfnamefont
  {T.}~\bibnamefont {Takabatake}},\ }\bibfield  {title} {\enquote {\bibinfo
  {title} {{Simultaneous superconducting and antiferroquadrupolar transitions
  in PrRh$_2$Zn$_{20}$}},}\ }\href@noop {} {\bibfield  {journal} {\bibinfo
  {journal} {Phys. Rev. B}\ }\textbf {\bibinfo {volume} {86}},\ \bibinfo
  {pages} {184426} (\bibinfo {year} {2012})}\BibitemShut {NoStop}%
\bibitem [{\citenamefont {Onimaru}\ \emph {et~al.}(2010)\citenamefont
  {Onimaru}, \citenamefont {T.~Matsumoto}, \citenamefont {F.~Inoue},
  \citenamefont {Umeo}, \citenamefont {Saiga}, \citenamefont {Matsushita},
  \citenamefont {Tamura}, \citenamefont {Nishimoto}, \citenamefont {Ishii},
  \citenamefont {Suzuki} \emph {et~al.}}]{onimaru2010superconductivity}%
  \BibitemOpen
  \bibfield  {author} {\bibinfo {author} {\bibfnamefont {Takahiro}\
  \bibnamefont {Onimaru}}, \bibinfo {author} {\bibfnamefont {Keisuke}\
  \bibnamefont {T.~Matsumoto}}, \bibinfo {author} {\bibfnamefont {Yukihiro}\
  \bibnamefont {F.~Inoue}}, \bibinfo {author} {\bibfnamefont {Kazunori}\
  \bibnamefont {Umeo}}, \bibinfo {author} {\bibfnamefont {Yuta}\ \bibnamefont
  {Saiga}}, \bibinfo {author} {\bibfnamefont {Yoshitaka}\ \bibnamefont
  {Matsushita}}, \bibinfo {author} {\bibfnamefont {Ryuji}\ \bibnamefont
  {Tamura}}, \bibinfo {author} {\bibfnamefont {Kazue}\ \bibnamefont
  {Nishimoto}}, \bibinfo {author} {\bibfnamefont {Isao}\ \bibnamefont {Ishii}},
  \bibinfo {author} {\bibfnamefont {Takashi}\ \bibnamefont {Suzuki}},  \emph
  {et~al.},\ }\bibfield  {title} {\enquote {\bibinfo {title}
  {{Superconductivity and structural phase transitions in caged compounds
  RT$_2$Zn$_{20}$ (R=La, Pr, T=Ru, Ir)}},}\ }\href@noop {} {\bibfield
  {journal} {\bibinfo  {journal} {J. Phys. Soc. Jpn.}\ }\textbf {\bibinfo
  {volume} {79}},\ \bibinfo {pages} {033704} (\bibinfo {year}
  {2010})}\BibitemShut {NoStop}%
\bibitem [{\citenamefont {Sakai}\ \emph {et~al.}(2012)\citenamefont {Sakai},
  \citenamefont {Kuga},\ and\ \citenamefont
  {Nakatsuji}}]{sakai2012superconductivity}%
  \BibitemOpen
  \bibfield  {author} {\bibinfo {author} {\bibfnamefont {Akito}\ \bibnamefont
  {Sakai}}, \bibinfo {author} {\bibfnamefont {Kentaro}\ \bibnamefont {Kuga}}, \
  and\ \bibinfo {author} {\bibfnamefont {Satoru}\ \bibnamefont {Nakatsuji}},\
  }\bibfield  {title} {\enquote {\bibinfo {title} {{Superconductivity in the
  ferroquadrupolar state in the quadrupolar Kondo lattice
  PrTi$_2$Al$_{20}$}},}\ }\href@noop {} {\bibfield  {journal} {\bibinfo
  {journal} {J. Phys. Soc. Jpn.}\ }\textbf {\bibinfo {volume} {81}},\ \bibinfo
  {pages} {083702} (\bibinfo {year} {2012})}\BibitemShut {NoStop}%
\bibitem [{\citenamefont {Matsubayashi}\ \emph {et~al.}(2012)\citenamefont
  {Matsubayashi}, \citenamefont {Tanaka}, \citenamefont {Sakai}, \citenamefont
  {Nakatsuji}, \citenamefont {Kubo},\ and\ \citenamefont
  {Uwatoko}}]{matsubayashi2012pressure}%
  \BibitemOpen
  \bibfield  {author} {\bibinfo {author} {\bibfnamefont {K.}~\bibnamefont
  {Matsubayashi}}, \bibinfo {author} {\bibfnamefont {T.}~\bibnamefont
  {Tanaka}}, \bibinfo {author} {\bibfnamefont {A.}~\bibnamefont {Sakai}},
  \bibinfo {author} {\bibfnamefont {S.}~\bibnamefont {Nakatsuji}}, \bibinfo
  {author} {\bibfnamefont {Y.}~\bibnamefont {Kubo}}, \ and\ \bibinfo {author}
  {\bibfnamefont {Y.}~\bibnamefont {Uwatoko}},\ }\bibfield  {title} {\enquote
  {\bibinfo {title} {{Pressure-induced heavy fermion superconductivity in the
  nonmagnetic quadrupolar system PrTi$_2$Al$_{20}$}},}\ }\href@noop {}
  {\bibfield  {journal} {\bibinfo  {journal} {Phys. Rev. Lett.}\ }\textbf
  {\bibinfo {volume} {109}},\ \bibinfo {pages} {187004} (\bibinfo {year}
  {2012})}\BibitemShut {NoStop}%
\bibitem [{\citenamefont {Matsubayashi}\ \emph {et~al.}(2014)\citenamefont
  {Matsubayashi}, \citenamefont {Tanaka}, \citenamefont {Suzuki}, \citenamefont
  {Sakai}, \citenamefont {Nakatsuji}, \citenamefont {Kitagawa}, \citenamefont
  {Kubo},\ and\ \citenamefont {Uwatoko}}]{matsubayashi2014heavy}%
  \BibitemOpen
  \bibfield  {author} {\bibinfo {author} {\bibfnamefont {Kazuyuki}\
  \bibnamefont {Matsubayashi}}, \bibinfo {author} {\bibfnamefont {Toshiki}\
  \bibnamefont {Tanaka}}, \bibinfo {author} {\bibfnamefont {Junichirou}\
  \bibnamefont {Suzuki}}, \bibinfo {author} {\bibfnamefont {Akito}\
  \bibnamefont {Sakai}}, \bibinfo {author} {\bibfnamefont {Satoru}\
  \bibnamefont {Nakatsuji}}, \bibinfo {author} {\bibfnamefont {Kentaro}\
  \bibnamefont {Kitagawa}}, \bibinfo {author} {\bibfnamefont {Yasunori}\
  \bibnamefont {Kubo}}, \ and\ \bibinfo {author} {\bibfnamefont {Yoshiya}\
  \bibnamefont {Uwatoko}},\ }\bibfield  {title} {\enquote {\bibinfo {title}
  {{Heavy Fermion Superconductivity under Pressure in the Quadrupole System
  PrTi$_2$Al$_{20}$}},}\ }in\ \href@noop {} {\emph {\bibinfo {booktitle}
  {Proceedings of the International Conference on Strongly Correlated Electron
  Systems (SCES2013)}}}\ (\bibinfo {year} {2014})\ p.\ \bibinfo {pages}
  {011077}\BibitemShut {NoStop}%
\bibitem [{\citenamefont {Tsujimoto}\ \emph {et~al.}(2014)\citenamefont
  {Tsujimoto}, \citenamefont {Matsumoto}, \citenamefont {Tomita}, \citenamefont
  {Sakai},\ and\ \citenamefont {Nakatsuji}}]{tsujimoto2014heavy}%
  \BibitemOpen
  \bibfield  {author} {\bibinfo {author} {\bibfnamefont {Masaki}\ \bibnamefont
  {Tsujimoto}}, \bibinfo {author} {\bibfnamefont {Yosuke}\ \bibnamefont
  {Matsumoto}}, \bibinfo {author} {\bibfnamefont {Takahiro}\ \bibnamefont
  {Tomita}}, \bibinfo {author} {\bibfnamefont {Akito}\ \bibnamefont {Sakai}}, \
  and\ \bibinfo {author} {\bibfnamefont {Satoru}\ \bibnamefont {Nakatsuji}},\
  }\bibfield  {title} {\enquote {\bibinfo {title} {{Heavy-fermion
  superconductivity in the quadrupole ordered state of PrV$_2$Al$_{20}$}},}\
  }\href@noop {} {\bibfield  {journal} {\bibinfo  {journal} {Phys. Rev. Lett.}\
  }\textbf {\bibinfo {volume} {113}},\ \bibinfo {pages} {267001} (\bibinfo
  {year} {2014})}\BibitemShut {NoStop}%
\bibitem [{\citenamefont {Iwasa}\ \emph {et~al.}(2017)\citenamefont {Iwasa},
  \citenamefont {Matsumoto}, \citenamefont {Onimaru}, \citenamefont
  {Takabatake}, \citenamefont {Mignot},\ and\ \citenamefont
  {Gukasov}}]{iwasa2017evidence}%
  \BibitemOpen
  \bibfield  {author} {\bibinfo {author} {\bibfnamefont {Kazuaki}\ \bibnamefont
  {Iwasa}}, \bibinfo {author} {\bibfnamefont {Keisuke~T.}\ \bibnamefont
  {Matsumoto}}, \bibinfo {author} {\bibfnamefont {Takahiro}\ \bibnamefont
  {Onimaru}}, \bibinfo {author} {\bibfnamefont {Toshiro}\ \bibnamefont
  {Takabatake}}, \bibinfo {author} {\bibfnamefont {Jean-Michel}\ \bibnamefont
  {Mignot}}, \ and\ \bibinfo {author} {\bibfnamefont {Arsen}\ \bibnamefont
  {Gukasov}},\ }\bibfield  {title} {\enquote {\bibinfo {title} {{Evidence for
  antiferromagnetic-type ordering of f-electron multipoles in
  PrIr$_2$Zn$_{20}$}},}\ }\href@noop {} {\bibfield  {journal} {\bibinfo
  {journal} {Phys. Rev. B}\ }\textbf {\bibinfo {volume} {95}},\ \bibinfo
  {pages} {155106} (\bibinfo {year} {2017})}\BibitemShut {NoStop}%
\bibitem [{\citenamefont {Taniguchi}\ \emph {et~al.}(2016)\citenamefont
  {Taniguchi}, \citenamefont {Yoshida}, \citenamefont {Takeda}, \citenamefont
  {Takigawa}, \citenamefont {Tsujimoto}, \citenamefont {Sakai}, \citenamefont
  {Matsumoto},\ and\ \citenamefont {Nakatsuji}}]{taniguchi2016nmr}%
  \BibitemOpen
  \bibfield  {author} {\bibinfo {author} {\bibfnamefont {Takanori}\
  \bibnamefont {Taniguchi}}, \bibinfo {author} {\bibfnamefont {Makoto}\
  \bibnamefont {Yoshida}}, \bibinfo {author} {\bibfnamefont {Hikaru}\
  \bibnamefont {Takeda}}, \bibinfo {author} {\bibfnamefont {Masashi}\
  \bibnamefont {Takigawa}}, \bibinfo {author} {\bibfnamefont {Masaki}\
  \bibnamefont {Tsujimoto}}, \bibinfo {author} {\bibfnamefont {Akito}\
  \bibnamefont {Sakai}}, \bibinfo {author} {\bibfnamefont {Yosuke}\
  \bibnamefont {Matsumoto}}, \ and\ \bibinfo {author} {\bibfnamefont {Satoru}\
  \bibnamefont {Nakatsuji}},\ }\bibfield  {title} {\enquote {\bibinfo {title}
  {{NMR Observation of Ferro-Quadrupole Order in PrTi$_2$Al$_{20}$}},}\
  }\href@noop {} {\bibfield  {journal} {\bibinfo  {journal} {J. Phys. Soc.
  Jpn.}\ }\textbf {\bibinfo {volume} {85}},\ \bibinfo {pages} {113703}
  (\bibinfo {year} {2016})}\BibitemShut {NoStop}%
\bibitem [{\citenamefont {Shiina}\ \emph {et~al.}(1997)\citenamefont {Shiina},
  \citenamefont {Shiba},\ and\ \citenamefont {Thalmeier}}]{shiina1997magnetic}%
  \BibitemOpen
  \bibfield  {author} {\bibinfo {author} {\bibfnamefont {Ryousuke}\
  \bibnamefont {Shiina}}, \bibinfo {author} {\bibfnamefont {Hiroyuki}\
  \bibnamefont {Shiba}}, \ and\ \bibinfo {author} {\bibfnamefont {Peter}\
  \bibnamefont {Thalmeier}},\ }\bibfield  {title} {\enquote {\bibinfo {title}
  {{Magnetic-field effects on quadrupolar ordering in a $\Gamma$ 8-quartet
  system CeB$_6$}},}\ }\href@noop {} {\bibfield  {journal} {\bibinfo  {journal}
  {J. Phys. Soc. Jpn.}\ }\textbf {\bibinfo {volume} {66}},\ \bibinfo {pages}
  {1741} (\bibinfo {year} {1997})}\BibitemShut {NoStop}%
\bibitem [{\citenamefont {Tsujimoto}\ \emph {et~al.}(2015)\citenamefont
  {Tsujimoto}, \citenamefont {Matsumoto},\ and\ \citenamefont
  {Nakatsuji}}]{tsujimoto2015anomalous}%
  \BibitemOpen
  \bibfield  {author} {\bibinfo {author} {\bibfnamefont {Masaki}\ \bibnamefont
  {Tsujimoto}}, \bibinfo {author} {\bibfnamefont {Yosuke}\ \bibnamefont
  {Matsumoto}}, \ and\ \bibinfo {author} {\bibfnamefont {Satoru}\ \bibnamefont
  {Nakatsuji}},\ }\bibfield  {title} {\enquote {\bibinfo {title} {{Anomalous
  specific heat behaviour in the quadrupolar Kondo system PrV$_2$Al$_{20}$}},}\
  }in\ \href@noop {} {\emph {\bibinfo {booktitle} {Journal of Physics:
  Conference Series}}},\ Vol.\ \bibinfo {volume} {592}\ (\bibinfo
  {organization} {IOP Publishing},\ \bibinfo {year} {2015})\ p.\ \bibinfo
  {pages} {012023}\BibitemShut {NoStop}%
\bibitem [{\citenamefont {L{\"o}hneysen}\ \emph {et~al.}(2007)\citenamefont
  {L{\"o}hneysen}, \citenamefont {Rosch}, \citenamefont {Vojta},\ and\
  \citenamefont {W{\"o}lfle}}]{lohneysen2007fermi}%
  \BibitemOpen
  \bibfield  {author} {\bibinfo {author} {\bibfnamefont {Hilbert~v}\
  \bibnamefont {L{\"o}hneysen}}, \bibinfo {author} {\bibfnamefont {Achim}\
  \bibnamefont {Rosch}}, \bibinfo {author} {\bibfnamefont {Matthias}\
  \bibnamefont {Vojta}}, \ and\ \bibinfo {author} {\bibfnamefont {Peter}\
  \bibnamefont {W{\"o}lfle}},\ }\bibfield  {title} {\enquote {\bibinfo {title}
  {Fermi-liquid instabilities at magnetic quantum phase transitions},}\
  }\href@noop {} {\bibfield  {journal} {\bibinfo  {journal} {Rev. Mod. Phys.}\
  }\textbf {\bibinfo {volume} {79}},\ \bibinfo {pages} {1015} (\bibinfo {year}
  {2007})}\BibitemShut {NoStop}%
\bibitem [{\citenamefont {de' Medici}\ \emph {et~al.}(2005)\citenamefont {de'
  Medici}, \citenamefont {Georges},\ and\ \citenamefont
  {Biermann}}]{de2005orbital}%
  \BibitemOpen
  \bibfield  {author} {\bibinfo {author} {\bibfnamefont {L}~\bibnamefont {de'
  Medici}}, \bibinfo {author} {\bibfnamefont {A.}~\bibnamefont {Georges}}, \
  and\ \bibinfo {author} {\bibfnamefont {S.}~\bibnamefont {Biermann}},\
  }\bibfield  {title} {\enquote {\bibinfo {title} {{Orbital-selective Mott
  transition in multiband systems: Slave-spin representation and dynamical
  mean-field theory}},}\ }\href@noop {} {\bibfield  {journal} {\bibinfo
  {journal} {Phys. Rev. B}\ }\textbf {\bibinfo {volume} {72}},\ \bibinfo
  {pages} {205124} (\bibinfo {year} {2005})}\BibitemShut {NoStop}%
\bibitem [{\citenamefont {de' Medici}\ \emph {et~al.}(2009)\citenamefont {de'
  Medici}, \citenamefont {Hassan}, \citenamefont {Capone},\ and\ \citenamefont
  {Dai}}]{de2009orbital}%
  \BibitemOpen
  \bibfield  {author} {\bibinfo {author} {\bibfnamefont {Luca}\ \bibnamefont
  {de' Medici}}, \bibinfo {author} {\bibfnamefont {S.R.}\ \bibnamefont
  {Hassan}}, \bibinfo {author} {\bibfnamefont {Massimo}\ \bibnamefont
  {Capone}}, \ and\ \bibinfo {author} {\bibfnamefont {Xi}~\bibnamefont {Dai}},\
  }\bibfield  {title} {\enquote {\bibinfo {title} {{Orbital-selective Mott
  transition out of band degeneracy lifting}},}\ }\href@noop {} {\bibfield
  {journal} {\bibinfo  {journal} {Phys. Rev. Lett.}\ }\textbf {\bibinfo
  {volume} {102}},\ \bibinfo {pages} {126401} (\bibinfo {year}
  {2009})}\BibitemShut {NoStop}%
\bibitem [{\citenamefont {Motrunich}(2005)}]{Motrunich2005}%
  \BibitemOpen
  \bibfield  {author} {\bibinfo {author} {\bibfnamefont {Olexei~I.}\
  \bibnamefont {Motrunich}},\ }\bibfield  {title} {\enquote {\bibinfo {title}
  {{Variational study of triangular lattice spin-$1/2$ model with ring
  exchanges and spin liquid state in
  $\ensuremath{\kappa}\text{\ensuremath{-}}{(\mathrm{ET})}_{2}{\mathrm{Cu}}_{2}{(\mathrm{CN})}_{3}$}},}\
  }\href {\doibase 10.1103/PhysRevB.72.045105} {\bibfield  {journal} {\bibinfo
  {journal} {Phys. Rev. B}\ }\textbf {\bibinfo {volume} {72}},\ \bibinfo
  {pages} {045105} (\bibinfo {year} {2005})}\BibitemShut {NoStop}%
\bibitem [{\citenamefont {Motrunich}\ and\ \citenamefont
  {Fisher}(2007)}]{Motrunich2007}%
  \BibitemOpen
  \bibfield  {author} {\bibinfo {author} {\bibfnamefont {Olexei~I.}\
  \bibnamefont {Motrunich}}\ and\ \bibinfo {author} {\bibfnamefont {Matthew
  P.~A.}\ \bibnamefont {Fisher}},\ }\bibfield  {title} {\enquote {\bibinfo
  {title} {{{$d$-wave} correlated critical Bose liquids in two dimensions}},}\
  }\href {\doibase 10.1103/PhysRevB.75.235116} {\bibfield  {journal} {\bibinfo
  {journal} {Phys. Rev. B}\ }\textbf {\bibinfo {volume} {75}},\ \bibinfo
  {pages} {235116} (\bibinfo {year} {2007})}\BibitemShut {NoStop}%
\bibitem [{\citenamefont {Sheng}\ \emph {et~al.}(2008)\citenamefont {Sheng},
  \citenamefont {Motrunich}, \citenamefont {Trebst}, \citenamefont {Gull},\
  and\ \citenamefont {Fisher}}]{sheng2008boson}%
  \BibitemOpen
  \bibfield  {author} {\bibinfo {author} {\bibfnamefont {D.~N.}\ \bibnamefont
  {Sheng}}, \bibinfo {author} {\bibfnamefont {Olexei~I.}\ \bibnamefont
  {Motrunich}}, \bibinfo {author} {\bibfnamefont {Simon}\ \bibnamefont
  {Trebst}}, \bibinfo {author} {\bibfnamefont {Emanuel}\ \bibnamefont {Gull}},
  \ and\ \bibinfo {author} {\bibfnamefont {Matthew P.~A.}\ \bibnamefont
  {Fisher}},\ }\bibfield  {title} {\enquote {\bibinfo {title} {{Strong-coupling
  phases of frustrated bosons on a two-leg ladder with ring exchange}},}\
  }\href {\doibase 10.1103/PhysRevB.78.054520} {\bibfield  {journal} {\bibinfo
  {journal} {Phys. Rev. B}\ }\textbf {\bibinfo {volume} {78}},\ \bibinfo
  {pages} {054520} (\bibinfo {year} {2008})}\BibitemShut {NoStop}%
\bibitem [{\citenamefont {Sheng}\ \emph {et~al.}(2009)\citenamefont {Sheng},
  \citenamefont {Motrunich},\ and\ \citenamefont {Fisher}}]{sheng2009spin}%
  \BibitemOpen
  \bibfield  {author} {\bibinfo {author} {\bibfnamefont {D.N.}\ \bibnamefont
  {Sheng}}, \bibinfo {author} {\bibfnamefont {Olexei~I.}\ \bibnamefont
  {Motrunich}}, \ and\ \bibinfo {author} {\bibfnamefont {Matthew~P.A.}\
  \bibnamefont {Fisher}},\ }\bibfield  {title} {\enquote {\bibinfo {title}
  {{Spin Bose-metal phase in a spin-1/2 model with ring exchange on a two-leg
  triangular strip}},}\ }\href@noop {} {\bibfield  {journal} {\bibinfo
  {journal} {Phys. Rev. B}\ }\textbf {\bibinfo {volume} {79}},\ \bibinfo
  {pages} {205112} (\bibinfo {year} {2009})}\BibitemShut {NoStop}%
\bibitem [{\citenamefont {Grover}\ \emph {et~al.}(2010)\citenamefont {Grover},
  \citenamefont {Trivedi}, \citenamefont {Senthil},\ and\ \citenamefont
  {Lee}}]{grover2010weak}%
  \BibitemOpen
  \bibfield  {author} {\bibinfo {author} {\bibfnamefont {Tarun}\ \bibnamefont
  {Grover}}, \bibinfo {author} {\bibfnamefont {N.}~\bibnamefont {Trivedi}},
  \bibinfo {author} {\bibfnamefont {T.}~\bibnamefont {Senthil}}, \ and\
  \bibinfo {author} {\bibfnamefont {Patrick~A.}\ \bibnamefont {Lee}},\
  }\bibfield  {title} {\enquote {\bibinfo {title} {{Weak Mott insulators on the
  triangular lattice: possibility of a gapless nematic quantum spin liquid}},}\
  }\href@noop {} {\bibfield  {journal} {\bibinfo  {journal} {Phys. Rev. B}\
  }\textbf {\bibinfo {volume} {81}},\ \bibinfo {pages} {245121} (\bibinfo
  {year} {2010})}\BibitemShut {NoStop}%
\bibitem [{\citenamefont {Mendive-Tapia}\ and\ \citenamefont
  {Staunton}(2017)}]{Staunton2017}%
  \BibitemOpen
  \bibfield  {author} {\bibinfo {author} {\bibfnamefont {Eduardo}\ \bibnamefont
  {Mendive-Tapia}}\ and\ \bibinfo {author} {\bibfnamefont {Julie~B.}\
  \bibnamefont {Staunton}},\ }\bibfield  {title} {\enquote {\bibinfo {title}
  {Theory of magnetic ordering in the heavy rare earths: Ab initio electronic
  origin of pair- and four-spin interactions},}\ }\href {\doibase
  10.1103/PhysRevLett.118.197202} {\bibfield  {journal} {\bibinfo  {journal}
  {Phys. Rev. Lett.}\ }\textbf {\bibinfo {volume} {118}},\ \bibinfo {pages}
  {197202} (\bibinfo {year} {2017})}\BibitemShut {NoStop}%
\bibitem [{\citenamefont {Akagi}\ \emph {et~al.}(2012)\citenamefont {Akagi},
  \citenamefont {Udagawa},\ and\ \citenamefont {Motome}}]{akagi2012hidden}%
  \BibitemOpen
  \bibfield  {author} {\bibinfo {author} {\bibfnamefont {Yutaka}\ \bibnamefont
  {Akagi}}, \bibinfo {author} {\bibfnamefont {Masafumi}\ \bibnamefont
  {Udagawa}}, \ and\ \bibinfo {author} {\bibfnamefont {Yukitoshi}\ \bibnamefont
  {Motome}},\ }\bibfield  {title} {\enquote {\bibinfo {title} {{Hidden
  multiple-spin interactions as an origin of spin scalar chiral order in
  frustrated Kondo lattice models}},}\ }\href@noop {} {\bibfield  {journal}
  {\bibinfo  {journal} {Phys. Rev. Lett.}\ }\textbf {\bibinfo {volume} {108}},\
  \bibinfo {pages} {096401} (\bibinfo {year} {2012})}\BibitemShut {NoStop}%
\bibitem [{\citenamefont {Hattori}\ and\ \citenamefont
  {Tsunetsugu}(2014)}]{hattori2014antiferro}%
  \BibitemOpen
  \bibfield  {author} {\bibinfo {author} {\bibfnamefont {Kazumasa}\
  \bibnamefont {Hattori}}\ and\ \bibinfo {author} {\bibfnamefont {Hirokazu}\
  \bibnamefont {Tsunetsugu}},\ }\bibfield  {title} {\enquote {\bibinfo {title}
  {{Antiferro Quadrupole Orders in Non-Kramers Doublet Systems}},}\ }\href@noop
  {} {\bibfield  {journal} {\bibinfo  {journal} {J. Phys. Soc. Jpn.}\ }\textbf
  {\bibinfo {volume} {83}},\ \bibinfo {pages} {034709} (\bibinfo {year}
  {2014})}\BibitemShut {NoStop}%
\bibitem [{\citenamefont {Stevens}(1952)}]{stevens1952matrix}%
  \BibitemOpen
  \bibfield  {author} {\bibinfo {author} {\bibfnamefont {K.W.H.}\ \bibnamefont
  {Stevens}},\ }\bibfield  {title} {\enquote {\bibinfo {title} {Matrix elements
  and operator equivalents connected with the magnetic properties of rare earth
  ions},}\ }\href@noop {} {\bibfield  {journal} {\bibinfo  {journal}
  {Proceedings of the Physical Society. Section A}\ }\textbf {\bibinfo {volume}
  {65}},\ \bibinfo {pages} {209} (\bibinfo {year} {1952})}\BibitemShut
  {NoStop}%
\bibitem [{\citenamefont {Lea}\ \emph {et~al.}(1962)\citenamefont {Lea},
  \citenamefont {Leask},\ and\ \citenamefont {Wolf}}]{lea1962raising}%
  \BibitemOpen
  \bibfield  {author} {\bibinfo {author} {\bibfnamefont {K.R.}\ \bibnamefont
  {Lea}}, \bibinfo {author} {\bibfnamefont {M.J.M.}\ \bibnamefont {Leask}}, \
  and\ \bibinfo {author} {\bibfnamefont {W.P.}\ \bibnamefont {Wolf}},\
  }\bibfield  {title} {\enquote {\bibinfo {title} {{The raising of angular
  momentum degeneracy of f-electron terms by cubic crystal fields}},}\
  }\href@noop {} {\bibfield  {journal} {\bibinfo  {journal} {Journal of Physics
  and Chemistry of Solids}\ }\textbf {\bibinfo {volume} {23}},\ \bibinfo
  {pages} {1381} (\bibinfo {year} {1962})}\BibitemShut {NoStop}%
\bibitem [{Note1()}]{Note1}%
  \BibitemOpen
  \bibinfo {note} {We note in passing that there are symmetry allowed
  three-spin interactions, but they are highly frustrated; these will be
  explored elsewhere.}\BibitemShut {Stop}%
\bibitem [{\citenamefont {Bergman}\ \emph {et~al.}(2007)\citenamefont
  {Bergman}, \citenamefont {Alicea}, \citenamefont {Gull}, \citenamefont
  {Trebst},\ and\ \citenamefont {Balents}}]{bergman2007order}%
  \BibitemOpen
  \bibfield  {author} {\bibinfo {author} {\bibfnamefont {Doron}\ \bibnamefont
  {Bergman}}, \bibinfo {author} {\bibfnamefont {Jason}\ \bibnamefont {Alicea}},
  \bibinfo {author} {\bibfnamefont {Emanuel}\ \bibnamefont {Gull}}, \bibinfo
  {author} {\bibfnamefont {Simon}\ \bibnamefont {Trebst}}, \ and\ \bibinfo
  {author} {\bibfnamefont {Leon}\ \bibnamefont {Balents}},\ }\bibfield  {title}
  {\enquote {\bibinfo {title} {Order-by-disorder and spiral spin-liquid in
  frustrated diamond-lattice antiferromagnets},}\ }\href@noop {} {\bibfield
  {journal} {\bibinfo  {journal} {Nature Physics}\ }\textbf {\bibinfo {volume}
  {3}},\ \bibinfo {pages} {487} (\bibinfo {year} {2007})}\BibitemShut {NoStop}%
\bibitem [{\citenamefont {Lee}\ and\ \citenamefont
  {Balents}(2008)}]{lee2008theory}%
  \BibitemOpen
  \bibfield  {author} {\bibinfo {author} {\bibfnamefont {SungBin}\ \bibnamefont
  {Lee}}\ and\ \bibinfo {author} {\bibfnamefont {Leon}\ \bibnamefont
  {Balents}},\ }\bibfield  {title} {\enquote {\bibinfo {title} {{Theory of the
  ordered phase in A-site antiferromagnetic spinels}},}\ }\href@noop {}
  {\bibfield  {journal} {\bibinfo  {journal} {Phys. Rev. B}\ }\textbf {\bibinfo
  {volume} {78}},\ \bibinfo {pages} {144417} (\bibinfo {year}
  {2008})}\BibitemShut {NoStop}%
\bibitem [{\citenamefont {Lee}\ \emph {et~al.}()\citenamefont {Lee},
  \citenamefont {Attig}, \citenamefont {Freyer}, \citenamefont {Paramekanti},
  \citenamefont {Trebst},\ and\ \citenamefont {Kim}}]{inpreparation}%
  \BibitemOpen
  \bibfield  {author} {\bibinfo {author} {\bibfnamefont {S.B.}\ \bibnamefont
  {Lee}}, \bibinfo {author} {\bibfnamefont {J.}~\bibnamefont {Attig}}, \bibinfo
  {author} {\bibfnamefont {F.}~\bibnamefont {Freyer}}, \bibinfo {author}
  {\bibfnamefont {A.}~\bibnamefont {Paramekanti}}, \bibinfo {author}
  {\bibfnamefont {S.}~\bibnamefont {Trebst}}, \ and\ \bibinfo {author}
  {\bibfnamefont {Y.B.}\ \bibnamefont {Kim}},\ }\href@noop {} {\bibinfo
  {journal} {in preparation}\ }\BibitemShut {NoStop}%
\end{thebibliography}%


\newpage

\appendix

\section{Supplementary numerical data}

\begin{figure}[h!]
	\centering
	\includegraphics[width=.9\columnwidth]{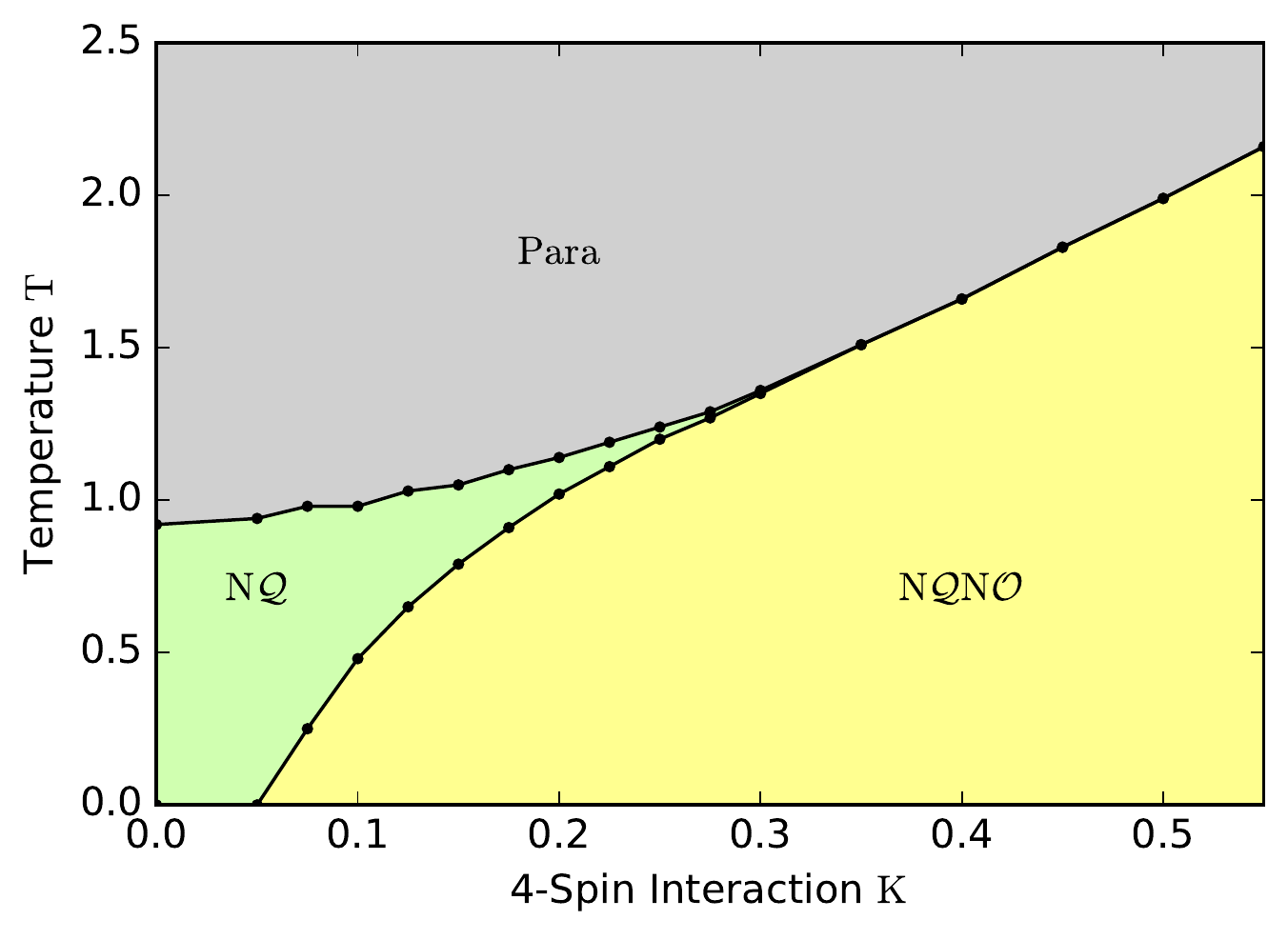}
	\caption{Finite-temperature phase diagram for a constant $J_2=0$ cut through the phase diagram of Fig.~\ref{fig:finiteT-phase-diag}.}
	\label{fig:J2_slice1}
\end{figure}

\begin{figure}[h!]
	\centering
	\includegraphics[width=.9\columnwidth]{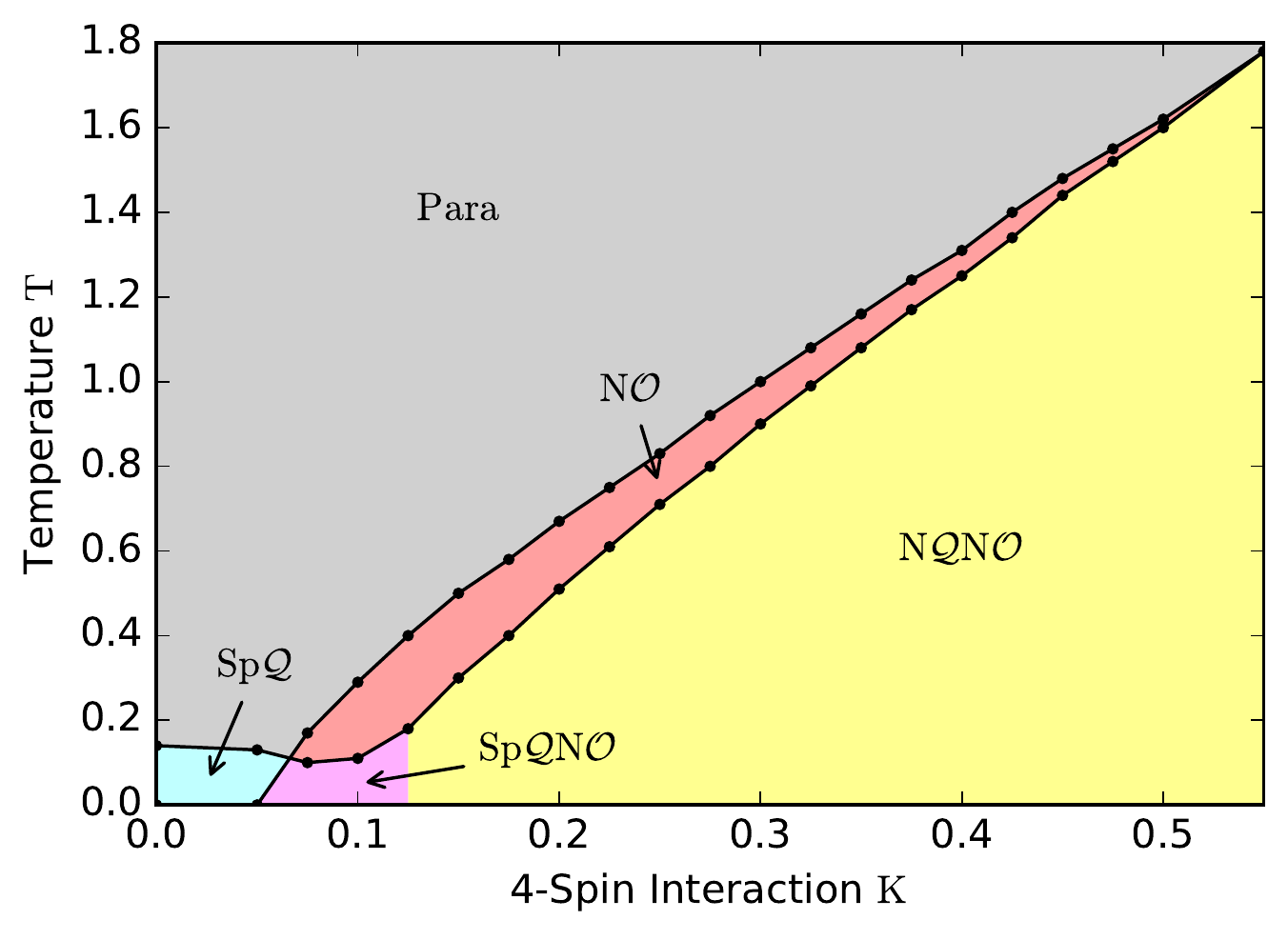}
	\caption{Finite-temperature phase diagram along a constant $J_2=0.25$ cut through the phase diagram of Fig.~\ref{fig:finiteT-phase-diag}.}
	\label{fig:J2_slice2}
\end{figure}

\begin{figure}[h]
	\centering
	\includegraphics[width=\columnwidth]{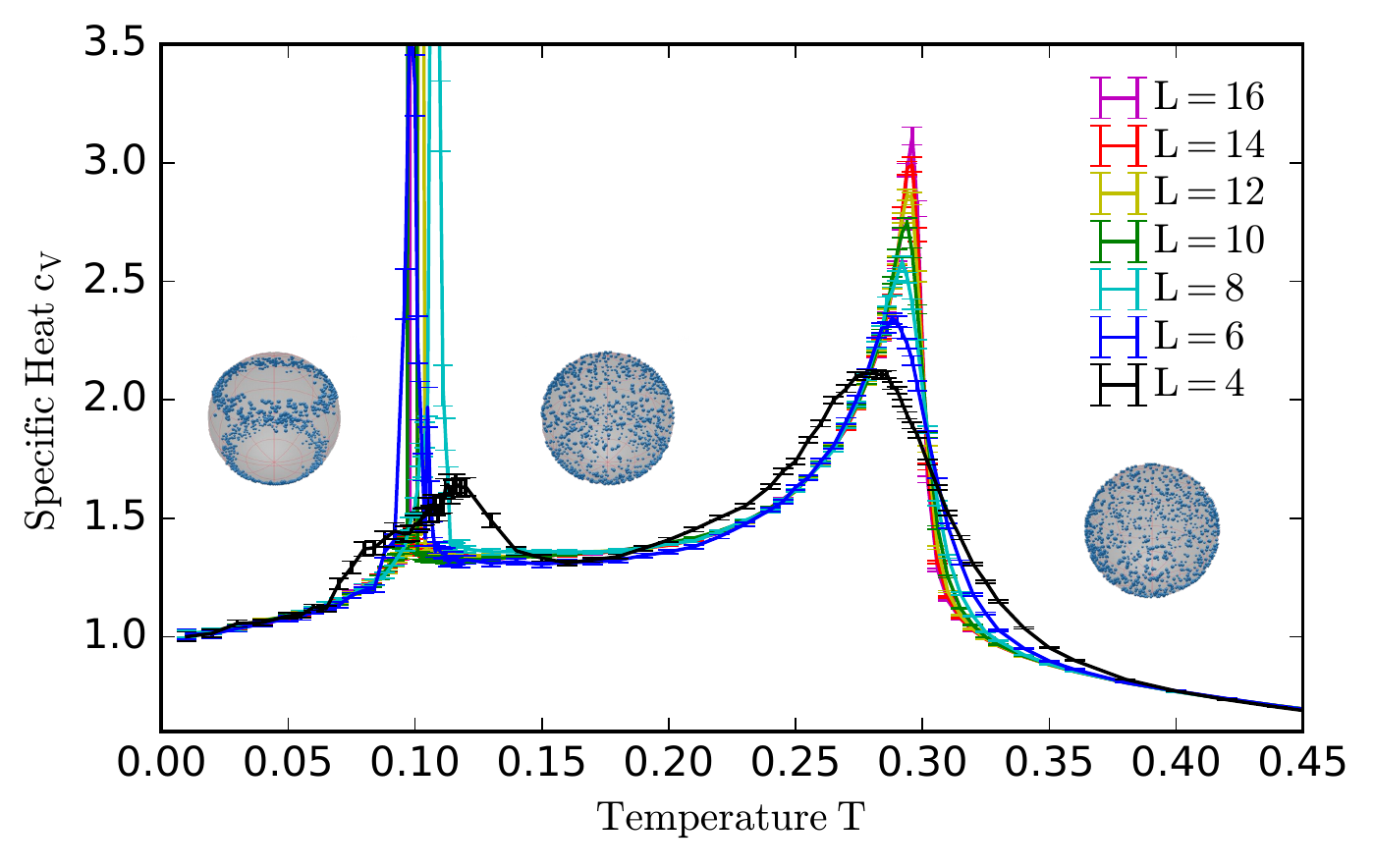}
	\caption{Double peak structure in specific heat scans for the two-stage ordering 
			from paramagnet to \nO\ to coexisting \spQ\nO\ order at 
		      zero temperature ($J_2 = 0.25$;  $K = -0.1$) for various system sizes.
	}
	\label{fig:CV1}
\end{figure}

\begin{figure}[h]
	\centering
	\includegraphics[width=\columnwidth]{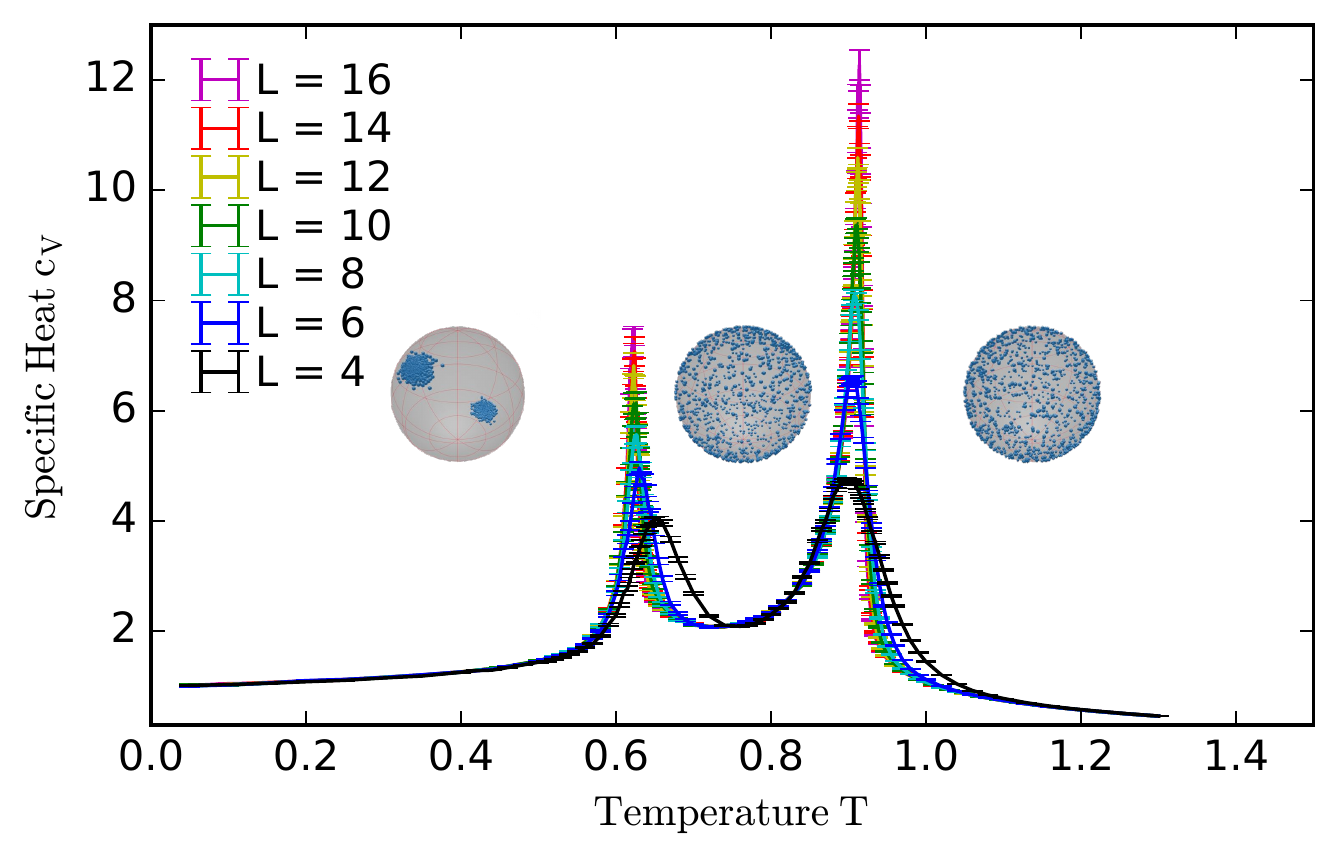}
	\caption{Double peak structure in specific heat scans for the two-stage ordering
	 	       from paramagnet to \nO\ to the coexisting \nQ\nO\ order at 
		      zero temperature ($J_2 = 0.35$;   $K = -0.3$) for various system sizes.
		}
	\label{fig:CV2}
\end{figure}

To round of our manuscript we present, in this appendix, some additional numerical data for the two-stage multipolar ordering transitions
in various parts of the phase diagram of the $J_1$-$J_2$-$K$ model presented in Fig.~\ref{fig:finiteT-phase-diag}.
This includes the finite-temperature phase diagrams of Figs.~\ref{fig:J2_slice1} and \ref{fig:J2_slice2} for constant $J_2=0$ and $J_2=0.25$ cuts through the phase diagram of Fig.~\ref{fig:finiteT-phase-diag}.
We further present, in Figs.~\ref{fig:CV1} and \ref{fig:CV2}, explicit numerical data for specific heat scans revealing the two-stage thermal transitions into low-temperature
spQ\nO\ and \nQ\nO\ orders, respectively.

\clearpage

\end{document}